\documentclass[journal]{IEEEtran}

\usepackage{xcolor,soul,framed} 
\colorlet{shadecolor}{yellow}
\usepackage{graphicx}
\graphicspath{{../pdf/}{../jpeg/}}
\DeclareGraphicsExtensions{.pdf,.jpeg,.png}
\usepackage[cmex10]{amsmath}
\usepackage{latexsym,bm,amssymb}
\usepackage{array}
\usepackage{mdwmath}
\usepackage{mdwtab}
\usepackage{eqparbox}
\usepackage{url}
\usepackage{enumerate}
\usepackage{amsfonts}
\usepackage{algorithmic}
\usepackage{algorithm}
\usepackage[numbers,sort&compress]{natbib}

\usepackage[justification=raggedright,singlelinecheck=false]{caption} 
\usepackage{mathrsfs}
\usepackage{diagbox}
\usepackage{subcaption}
\usepackage[justification=raggedright,singlelinecheck=false]{caption} 
\usepackage{tabularx} 

\usepackage{multirow}

\usepackage{booktabs} 

\usepackage{footnote}

\usepackage{nomencl}
\makenomenclature
\usepackage{ifthen}
  \renewcommand{\nomgroup}[1]{%
  \item[\bfseries
  \ifthenelse{\equal{#1}{A}}{Symbols of UC}{%
  \ifthenelse{\equal{#1}{S}}{Symbols of CL}{}}%
  ]}

\renewcommand{\baselinestretch}{1}

\begin{document}

\captionsetup{font={small}}

\title{Power Swing Trajectory Influenced by Virtual Impedance-Based Current-Limiting Strategy}

\author{Yanshu Niu,~\IEEEmembership{Student Member,~IEEE,} Zhe Yang,~\IEEEmembership{Member,~IEEE,} and Bikash C. Pal,~\IEEEmembership{Fellow,~IEEE}


}


\maketitle

\begin{abstract}
Grid-forming (GFM) inverter-based resources (IBRs) can emulate the external characteristics of synchronous generators (SGs) through appropriate control loop design. However, in systems with GFM IBRs, the apparent impedance trajectory under current limitation differs significantly from that of SG-based systems due to the limited overcurrent capability of power electronic devices. This difference challenges the power swing detection functions of distance relays designed for SG-based systems. This paper presents a theoretical analysis of the apparent impedance trajectory over a full power swing cycle under two typical current-limiting strategies: variable virtual impedance (VI) and adaptive VI. The analysis reveals that the trajectory under VI current-limiting strategies differs significantly from that of a conventional SG. The results also indicate that the control parameters affect the characteristics of the trajectory. In addition, the new trajectories challenge conventional power swing detection functions, increasing the risk of malfunction. Furthermore, the implementation of VI leads to a deterioration in system stability. The theoretical analysis is further validated through simulations on the MATLAB/Simulink platform.
\end{abstract}

\begin{IEEEkeywords}
power swing detection, virtual impedance, current limitation, grid-forming control, apparent impedance trajectory
\end{IEEEkeywords}





\section{Introduction}\label{introduction}
\label{SectionI_Introduction}
\IEEEPARstart{D}{riven} by environmental sustainability considerations, an increasing number of inverter-based resources (IBRs) powered by renewable energy are replacing conventional fossil-fuel-based synchronous generators (SGs) in power systems. As the penetration of IBRs increases, grid-forming (GFM) IBRs offer advantages over grid-following (GFL) IBRs due to their ability to provide essential services, such as frequency regulation and voltage support~\cite{li2022revisiting,lasseter2020}. As a result, GFM IBRs are expected to play a crucial role in future power systems. However, the unique control-based external characteristics of GFM IBRs under current limitation pose challenges to legacy protection systems, with power swing detection being a key issue~\cite{munz2024,haddadi2017}.

The coordination between the two power swing detection functions, power swing blocking (PSB) and out-of-step tripping (OST), aims to block distance protection during stable power swings and release blocking during unstable power swings~\cite{nerc2013,fischer2012tutorial,psrc2005}. The effectiveness of power swing detection stems from the well-characterised apparent impedance trajectory in SGs-based systems. This trajectory can correctly reflect the power swing angle. The inherent rigid-body inertia of the SGs makes the rate of change of the trajectory during power swing significantly slower than that during faults events. The power swing detection functions in modern commercial distance relays are still designed for SG-based power systems~\cite{ge2023,rel670manual,siemens2023}. However, although the external characteristics of SGs can be emulated by GFM IBRs, their behaviour is governed by control loops, making their external characteristics dependent on control parameters. Moreover, as the power electronic devices in IBRs have limited overcurrent capacity, virtual impedance (VI) or current saturation algorithms are typically employed to limit overcurrent~\cite{fan2022,kkuni2023,qoria2020}. This paper focuses on GFM IBRs operating under the VI current-limiting strategy. The external characteristics of GFM IBRs in the current limiting mode result in a power swing trajectory distinct from that of SGs, posing challenges for power swing detection.
\IEEEpubidadjcol

Several studies have reported the impact of IBRs on legacy distance protection systems~\cite{liu2022,baeckeland2022,cao2023,cao2024,johansson2024}, focusing primarily on fault events, while their effect on power swing detection functions remains underexplored. In~\cite{Rao2022}, a protection scheme is proposed to distinguish between symmetrical faults and symmetrical power swings; however, its applicability is limited to Type-III wind turbine generators (WTG) and is not generally applicable to IBRs. In~\cite{Haddadi2019}, the power swing apparent impedance trajectory of Type-III WTGs is investigated, and the research in~\cite{Haddadi2021} further extends the scope to power systems with high penetration of IBRs. However, these studies are primarily qualitative and descriptive, relying on simulation-based case studies rather than theoretical analysis to uncover the underlying mechanisms. A limited number of studies have theoretically investigated the impact of IBRs on protection systems during power swing, analyzing their dynamic characteristics from the perspective of control loops~\cite{xiong2023,xiong2024,nasr2024part1,nasr2024part2}. In~\cite{xiong2023} and~\cite{xiong2024}, the authors conduct a detailed analysis of the apparent impedance trajectory under current limitation, with an emphasis on the control mechanism of the phase-locked loop (PLL). In~\cite{nasr2024part1} and~\cite{nasr2024part2}, after developing a dynamic model, the authors conclude that the DC-link voltage control dynamics of IBRs can significantly amplify the rate of impedance change, posing a potential risk of relay malfunctions. While these studies focus on GFL IBRs, GFM IBRs remain unexamined. The mechanism by which the resynchronization process after fault clearance affects the power swing detection functions in GFM IBR-based systems has been discussed in~\cite{Xiong2023unlimited}, whereas the role of current limitation was not considered. The impact of the current saturation algorithm on the power swing trajectory has been analyzed in~\cite{niu2025impedance}; however, VI is not within its scope. To the best of the authors' knowledge, no prior research has investigated the impact of GFM IBRs under the VI current-limiting strategy on the effectiveness of power swing detection.

To address this research gap, this paper theoretically analyses the apparent impedance trajectories over a full power swing cycle observed at the terminal of a GFM IBR under the VI current-limiting strategies. The impact of control parameters on the trajectories' behaviour is explored. Additionally, the distinct characteristics of the trajectories are shown to affect the performance of legacy power swing detection functions. The main contributions of this paper are as follows:

\begin{itemize}
    \item By analysing full-cycle power swing trajectories for both variable and adaptive VI current-limiting strategies, this study demonstrates that the apparent impedance trajectories differ fundamentally between these two strategies and from that of a conventional SG. 
    
    \item Theoretical analysis and simulation validation reveal that the apparent impedance trajectories under the VI current-limiting strategies can cause malfunctions in both PSB and OST, posing a potential risk to the proper operation of the protection system.
    
    \item It is verified by simulation analysis that the trajectories are affected by control parameters, which threaten the proper functioning of power swing detection. Additionally, the introduction of VI deteriorates the stability of the system.
\end{itemize}

The remainder of this paper is structured as follows. Section~\ref{SectionII_Current-Limiting Strategies for Grid-Forming Inverter Based on Virtual Impedance} presents the principles of variable and adaptive VI current-limiting strategies. In Section~\ref{SectionIII_Analysis of Power Swing Trajectories}, the full-cycle power swing trajectories under VI current-limiting strategies are theoretically analysed. Section~\ref{SectionIV_Impact of Power Swing Trajectories on Power Swing Detection} discusses the impact of power swing trajectories on power swing detection. Simulation results and case studies are provided in Section~\ref{SectionV_Simulation and Case Studies}. Finally, Section~\ref{SectionVI_Conclusion} concludes the paper.
\section{Current-Limiting Strategies for Grid-Forming Inverter Based on Virtual Impedance}
\label{SectionII_Current-Limiting Strategies for Grid-Forming Inverter Based on Virtual Impedance}
This section presents the grid-connected IBR system model and a detailed description of the GFM control loops, followed by an introduction to two typical VI-based current-limiting strategies.
\subsection{System Model}
\begin{figure*}[htbp]
    \centering
    \begin{subfigure}{0.8\textwidth}
        \centering
        \includegraphics[width=\textwidth, clip]{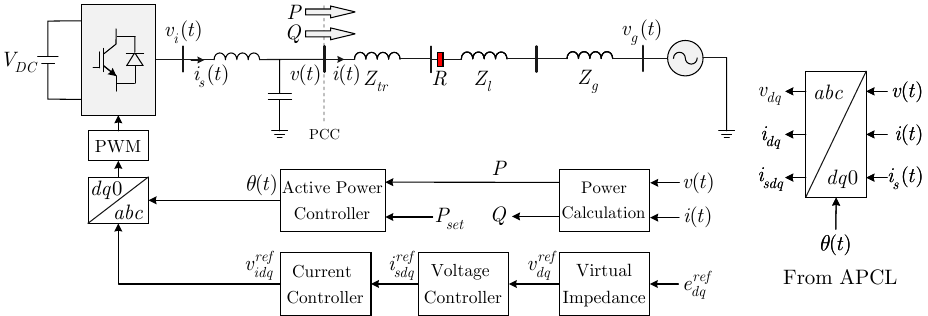} 
        \captionsetup{justification=centering} 
        \caption{}
    \end{subfigure}
    \begin{subfigure}{0.8\textwidth}
        \centering
        \includegraphics[width=\textwidth, clip]{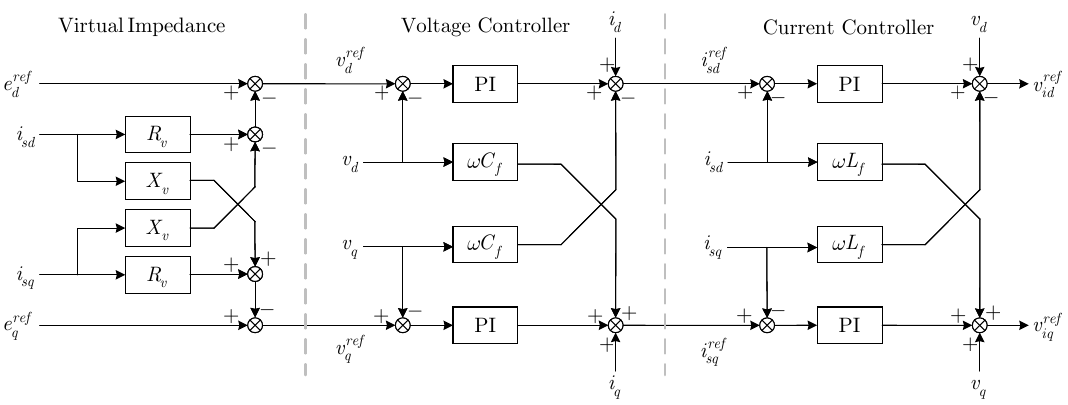} 
        \captionsetup{justification=centering} 
        \caption{}
    \end{subfigure}
    \caption{Grid-connected GFM IBR system. (a) System model and controller structure. (b) Cascaded control loops block diagram.}
    \label{fig:System Model}
\end{figure*}

The grid-connected GFM IBR system is shown in Fig.~\ref{fig:System Model}. In the system model of Fig.~\ref{fig:System Model}(a), the IBR is connected to the grid, which is modelled as a Thevenin equivalent voltage source \(\dot{V_g}\) in series with the impedance \(Z_g\), through an LC filter, a step-up transformer \(Z_{tr}\) and the power line \(Z_{l}\). Since the capacitance of the LC filter is small and not the focus of this paper, its impact is neglected in the subsequent analysis to simplify the investigation. The low-voltage side of the transformer is the point of common coupling (PCC). The total equivalent impedance between the PCC bus and the grid terminal is \(Z_{tr}+Z_{l}+Z_{g}=Z_{\Sigma}\angle \phi = R_{\Sigma}+jX_{\Sigma} \). In Fig.~\ref{fig:System Model}(a), a distance relay $R$ is installed near the IBR terminal of the power line, and is equipped with both conventional fault detection and power swing detection functions.
A typical control structure for GFM IBRs is the voltage-current dual-loop control, shown in Fig.~\ref{fig:System Model}(b); additionally, to mimic the rotor motion of SGs for generating the phase angle, this paper adopts an active power control loop (APCL) cascaded voltage-current dual-loop control scheme~\cite{li2023improved}. The input to the voltage control loop, defined as the setpoint for the IBR output voltage, is determined by the reactive power control loop (RPCL). However, since this is not the focus of this paper, the RPCL is neglected. Instead, the output voltage setpoint is assumed to be constant as \( \dot{{E}_{\text{ref}}} = e_{d}^{\text{ref}} + je_{q}^{\text{ref}}=1+j0 \). When the magnitude of the IBR's output current remains below the overcurrent threshold \( I_{\text{th}} \), the VI remains inactive, and the voltage reference at the PCC is \( \dot{{V}_{\text{ref}}} = \dot{E_{\text{ref}}}\). Conversely, once the current exceeds \( I_{\text{th}} \), the VI is activated, and the VI loop in Fig.~\ref{fig:System Model}(b) updates the reference to \( \dot{V_{\text{ref}}} \) for the voltage controller. The value of VI is determined by the VI current-limiting strategies.
\subsection{Variable Virtual Impedance}
The VI-based current-limiting strategy aims to limit current by adjusting the voltage reference of the PCC. When the current magnitude exceeds the current threshold \(I_{\text{th}}\), additional VI, \(Z_{\text{VI}}=R_{\text{VI}}+jX_{\text{VI}}\), is activated. The \(R_{\text{VI}}\) and \(X_{\text{VI}}\) satisfy the relationship
\begin{align}
\label{eq:VI Rv}
R_{\text{VI}} & =
\begin{cases} 
k_{\text{VI}}\left( I - I_{\text{th}} \right), & \text{if } I > I_{\text{th}}, \\
0, & \text{if } I \leq I_{\text{th}},
\end{cases} \\
\label{eq:VI Xv}
X_{\text{VI}} & = \alpha_{\text{VI}}R_{\text{VI}},
\end{align}
where \(I= \sqrt{i_{sd}^2+i_{sq}^2}\) represents the magnitude of the current; \(k_{\text{VI}}\) and \(\alpha_{\text{VI}}\) are defined as VI proportional gain and VI ratio, respectively. The dq-axis voltage drop across the VI is
\begin{align}
\label{eq:d-axis voltage drop on VI}
V_{d\text{VI}} & = R_{\text{VI}} i_{sd} - X_{\text{VI}} i_{sq}, \\
\label{eq:q-axis voltage drop on VI}
V_{q\text{VI}} & = R_{\text{VI}} i_{sq} + X_{\text{VI}} i_{sd}.
\end{align}
Thus, as shown in the VI loop of Fig.~\ref{fig:System Model}(b), the voltage references at the PCC are updated as follows:
\begin{align}
\label{eq:Updated Vdref Variable}
v_{d}^{\text{ref}} & = e_{d}^{\text{ref}} - V_{d\text{VI}}, \\
\label{eq:Updated Vqref Variable}
v_{q}^{\text{ref}} & = e_{q}^{\text{ref}} - V_{q\text{VI}}.
\end{align} 
In the aforementioned VI expressions~(\ref{eq:VI Rv}) and~(\ref{eq:VI Xv}), two key parameters are included: \(k_{\text{VI}}\), and \(\alpha_{\text{VI}}\). The ratio \(\alpha_{\text{VI}}\) determines the VI angle, set as \(\angle \phi\) in this paper, to match the angle of the total equivalent impedance between the PCC and the grid voltage source. The VI proportional gain \(k_{\text{VI}}\) determines the value of the VI, which is detailed as follows~\cite{paquette2014virtual}: 
\begin{equation}
k_{\text{VI}}=\frac{\Delta V}{\left( I_{\text{max}} - I_{\text{th}} \right)I_{\text{max}}\sqrt{1+\alpha_{\text{VI}}^2}},
\label{eq:General k_VI}
\end{equation}
where \(\Delta V\) represents the magnitude of the voltage drop across the VI. The detailed derivation of~(\ref{eq:General k_VI}) can be found in Appendix A.

The variable VI current-limiting strategy focuses on the protection perspective to prevent IBR damage from overcurrent during fault events. Specifically, under current limitation, the VI satisfies
\begin{equation}
\label{eq:Variable k_VI setting}
| \dot{E_{\text{ref}}} | = I_{\text{max}}|Z_{\text{VI}} | = I_{\text{max}}\sqrt{\left( 1+\alpha_{\text{VI}}^2 \right) k_{\text{VI}}^2\left( I_{\text{max}}-I_{\text{th}} \right)^2},
\end{equation}
to ensure that the overcurrent at the PCC during the most severe bolted three-phase short circuit events is limited to its maximum allowable current \(I_{\text{max}}\). If \(I>I_{\text{th}}\), the VI is activated with a constant proportional gain
\begin{equation}
\label{eq:Variable k_VI}
k_{\text{VI}}=\frac{| \dot{E_{\text{ref}}} |}{\left( I_{\text{max}} - I_{\text{th}} \right)I_{\text{max}}\sqrt{1+\alpha_{\text{VI}}^2}}.
\end{equation}
It is worth noting that even after setting the VI according to the principle aimed at protecting the IBR from overcurrent damage during the most severe short-circuit fault events, the IBR still has a risk of experiencing overcurrent-induced damage during power swing events.
\subsection{Adaptive Virtual Impedance}
\begin{figure}[htbp]
\centering
\includegraphics[width=2.6in]{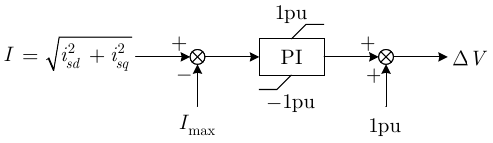}
\caption{Control block diagram in adaptive VI strategy to obtain the voltage drop \(\Delta V\) across the VI.}
\label{fig:Delta V}
\end{figure}

The adaptive VI-based current-limiting strategy is designed to limit overcurrent to an appropriate level under all conditions. The objective is achieved by dynamically adjusting the parameter \(k_{\text{VI}}\), which is no longer constant. The \(k_{\text{VI}}\) is determined in the same manner by the voltage drop across the VI in~(\ref{eq:General k_VI}). However, the voltage drop is not necessarily constant and is obtained through a PI control loop \cite{nasr2023}, as shown in Fig.~\ref{fig:Delta V}.
Within the PI controller block, the clamping anti-windup technique is employed to prevent integrator windup when the current magnitude remains below \(I_{\text{max}}\). The integrator loop in Fig.~\ref{fig:Delta V} ensures that \(\Delta V\) has a non-zero output only when \(I>I_{\text{max}}\), thereby activating the VI in the control loops.
\section{Analysis of Power Swing Trajectories}
\label{SectionIII_Analysis of Power Swing Trajectories}
This section considers three scenarios: the absence of current limitation, the variable VI strategy, and the adaptive VI strategy. The apparent impedance trajectory at the IBR terminal is analysed over a full power swing cycle.
\subsection{Trajectory in the Absence of Current Limitation}
\begin{figure}[htbp]
\centering
\includegraphics[width=2.6in]{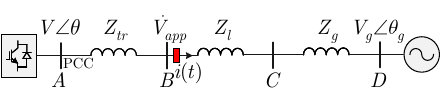}
\caption{Single-machine grid-connected GFM IBR system.}
\label{fig:Single Machine Grid Connected Circuit Model}
\end{figure}

\begin{figure}[htbp]
\centering
\includegraphics[width=0.98\columnwidth]{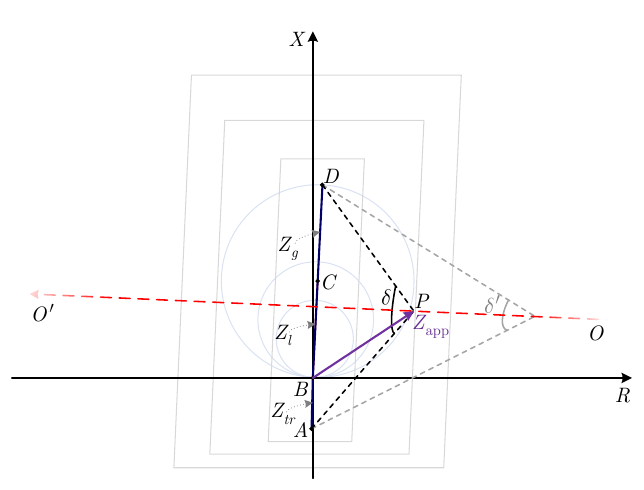}
\caption{Power swing trajectory in the absence of current limitation and for an SG.}
\label{fig:Unsat Trajectory}
\end{figure}

In the absence of current limitation, the analysis is conducted using a single machine grid-connected GFM IBR system in Fig.~\ref{fig:Single Machine Grid Connected Circuit Model}, which represents a simplified equivalent circuit of Fig.~\ref{fig:System Model}(a). The angle between the GFM IBR and the power grid is defined as the power angle \(\delta := \theta - \theta_{g}\). The IBR uses its local voltage output as the d-axis reference, i.e., \(\dot{V_{\text{ref}}} = 1 + j0\), which implies that \(\theta = 0\). Therefore, \(\theta_{g} = -\delta\).
For simplicity, the voltage magnitude at the GFM IBR output terminal, also known as the PCC, is assumed to be equal to the grid voltage magnitude, i.e., \(|\dot{V}_{g} | = |\dot{V} |\). The apparent impedance observed by the relay near Bus B in Fig.~\ref{fig:Single Machine Grid Connected Circuit Model} is calculated as
\begin{equation}
\label{eq:Apparent Impedance Unsaturated}
Z_{\text{app}} = \left( Z_{g} + Z_{l} - \frac{1}{2}Z_{\Sigma} \right) - j\frac{1}{2}Z_{\Sigma}\cot\frac{\delta}{2},
\end{equation}
where \(Z_{\Sigma}=Z_{tr} + Z_{l} + Z_{g}\), which is consistent with the derivation of apparent impedance during power swings in conventional SGs \cite{kundur1994}. The apparent impedance trajectory given by~(\ref{eq:Apparent Impedance Unsaturated}) is shown in Fig.~\ref{fig:Unsat Trajectory}.
In Fig.~\ref{fig:Unsat Trajectory}, point \(B\) is the location where the relay is installed. The red dashed line represents the trajectory. The angle \(\angle DPA\), denoted as \(\delta\), is the power angle between the IBR and the grid. The purple vector \(Z_{\text{app}}\) indicates the apparent impedance detected by the relay. During a full cycle, as the power angle varies from \(0\) to \(2\pi\), the apparent impedance moves from \(O\) to \(O'\), starting from positive infinity near the R-axis in the first quadrant and ending at negative infinity near the R-axis in the second quadrant. Each position on the impedance trajectory corresponds to a unique power angle \(\delta\).
\subsection{Trajectory with the Variable Virtual Impedance Strategy}
\begin{figure}[htbp]
\centering
\includegraphics[width=3.2 in]{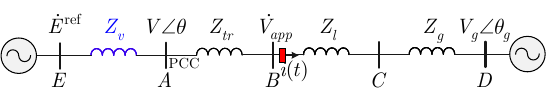}
\caption{Equivalent single-machine grid-connected GFM IBR system after implementing the VI current-limiting strategy.}
\label{fig:Single Machine Grid Connected Model with VI}
\end{figure}

\begin{figure}[htbp]
\centering
\includegraphics[width=0.98\columnwidth]{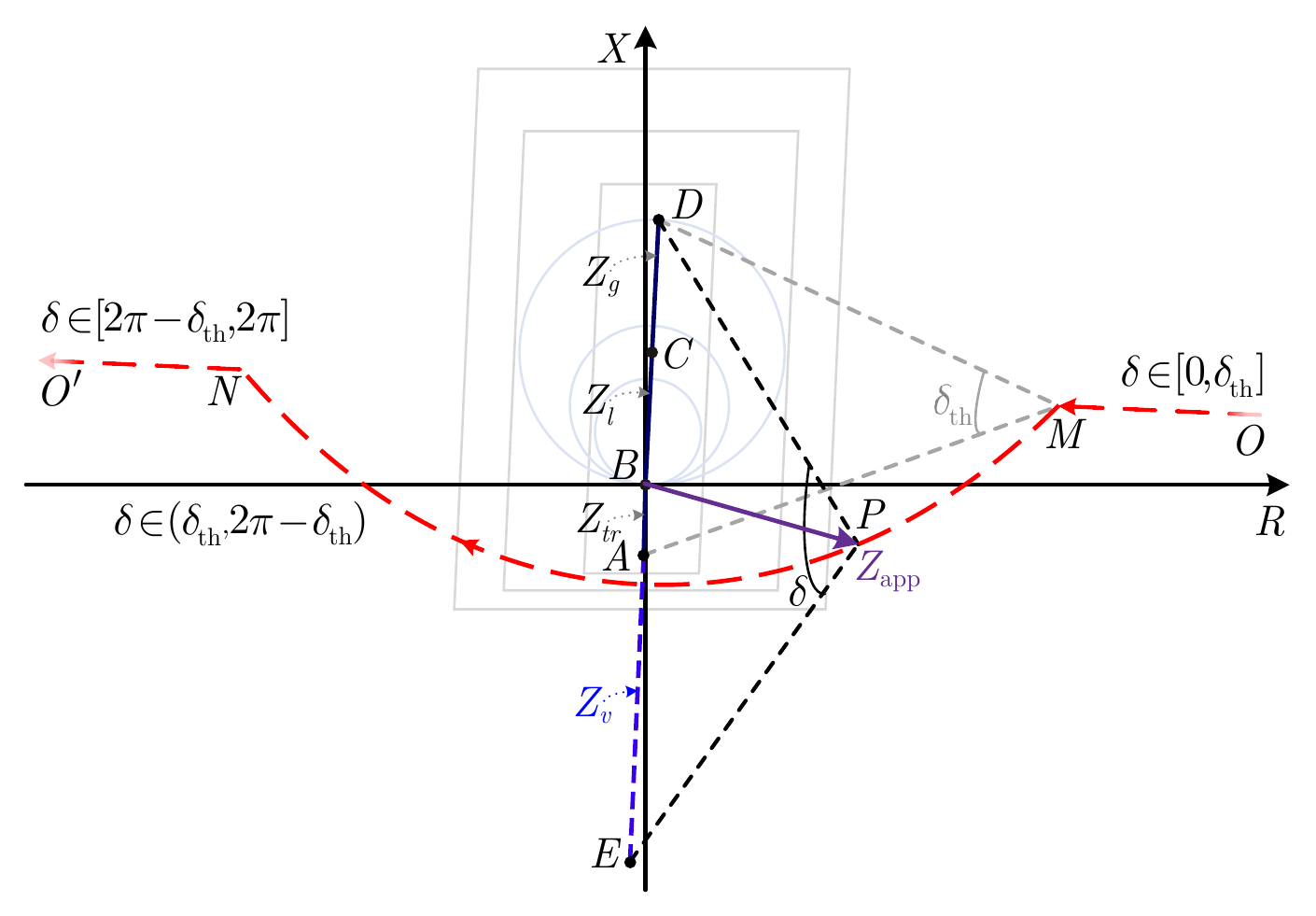}
\caption{Power swing trajectory with the variable VI current-limiting strategy}
\label{fig:Saturated Trajectory Variable}
\end{figure}

The output current of the GFM IBR varies with the power angle during power swings. Under the variable VI strategy, the VI is activated, if 
\begin{equation}
\label{eq:Virtual Impedance Activate Condition}
(i_{sd})^2+(i_{sq})^2 >I_{\text{th}}^2 .
\end{equation}
Otherwise, the VI remains inactive and is set to 0. The power angle \(\delta\) that causes the current to enter the current limiting mode satisfies
\begin{equation}
\label{eq:VI Activate Angle Condition}
\cos\delta < \frac{|\dot{{E}_{\text{ref}}}|^{2}+ | \dot{V_{g}} |^{2}-\left(| Z_{\Sigma} | I_{\text{th}} \right)^2}{2|\dot{{E}_{\text{ref}}} || \dot{V_{g}} |}.
\end{equation}
The critical power angle for the activation of VI is defined as:
\begin{equation}
\label{eq:VI Activate Angle}
\delta_{\text{th}} := \arccos \left[ \frac{| \dot{{E}_{\text{ref}}} |^{2}+| \dot{V_{g}} |^{2}-\left(| Z_{\Sigma} | I_{\text{th}} \right)^2}{2| \dot{{E}_{\text{ref}}} || \dot{V_{g}} |} \right].
\end{equation}
Therefore, during a full power swing cycle, the power angle sets for conditions where the variable VI is not activated and where it is activated are defined as follows, respectively.
\begin{align}
\mathcal{D}_{\text{unact}} &= \{ \delta \mid \delta \in [0, \delta_{\text{th}}] \cup [2\pi - \delta_{\text{th}}, 2\pi] \} 
\label{eq:Unactivate_Set} \\
\mathcal{D}_{\text{act}} &= \{ \delta \mid \delta \in (\delta_{\text{th}}, 2\pi - \delta_{\text{th}}) \} \label{eq:Activate_Set}
\end{align}
When the system operates in current limitation mode with the VI activated, the equivalent circuit is shown in Fig.~\ref{fig:Single Machine Grid Connected Model with VI}.
In this case, the voltage at the PCC can no longer track the voltage setpoint \(\dot{E}_{\text{ref}}\). Instead, \(\dot{E}_{\text{ref}}\) is used to generate a new PCC voltage reference through the voltage drop across the VI. Considering the effect of the VI, the current flowing through the system is
\begin{equation}\dot{I}=\frac{E_{\text{ref}}\angle0 - V_{g}\angle-\delta}{Z_{\Sigma}\angle\phi + \left[ k_{\text{VI}}( | \dot{I} | - I_{\text{th}})+j\alpha_{\text{VI}}k_{\text{VI}}( | \dot{I} | - I_{\text{th}})\right]}.
\end{equation}
Therefore, the apparent impedance observed by the relay near Bus B is calculated as
\begin{equation}
\label{eq:Apparent Impedance Saturated Variable}
Z_{\text{app}} = \left( Z_{g} + Z_{l}\right) + \frac{| \dot{V_{g}} |}{|\dot{I}|}\angle \left(-\delta-\theta_{i}\right),
\end{equation}
where \(\theta_{i}\) is the phase angle of the current \(\dot{I}\). This apparent impedance varies with the power angle, and its trajectory on the impedance plane is illustrated in Fig.~\ref{fig:Saturated Trajectory Variable}. When the power angle belongs to the set \(\mathcal{D}_{\text{unact}}\), i.e., \(\delta \in [0, \delta_{\text{th}}] \cup [2\pi - \delta_{\text{th}}, 2\pi]\), the corresponding trajectory segments in Fig.~\ref{fig:Saturated Trajectory Variable} are \(OM\) and \(NO'\), respectively. In this case, the current remains below the threshold \(I_{\text{th}}\), and the VI is not activated. Thus, the trajectory matches the case analysed in Section III.A. When the power angle belongs to the set \(\mathcal{D}_{\text{act}}\), the VI is activated, therefore, (\ref{eq:Apparent Impedance Saturated Variable}) is applied. The corresponding trajectory of this set in Fig.~\ref{fig:Saturated Trajectory Variable} is the red dashed curve \(MN\). The vector \(BD\) represents the impedance \(\left(Z_{g} + Z_{l}\right)\) as described in (\ref{eq:Apparent Impedance Saturated Variable}). The segment \(DP\) indicates the value of \(| \dot{V_{g}} | / | \dot{I} |\) at the angle \(\delta\). Since the current is not constant during the activation of the VI, the length of \(DP\) varies with the power angle \(\delta\).
\subsection{Trajectory with the Adaptive Virtual Impedance Strategy}
\begin{figure}[htbp]
\centering
\includegraphics[width=0.98\columnwidth]{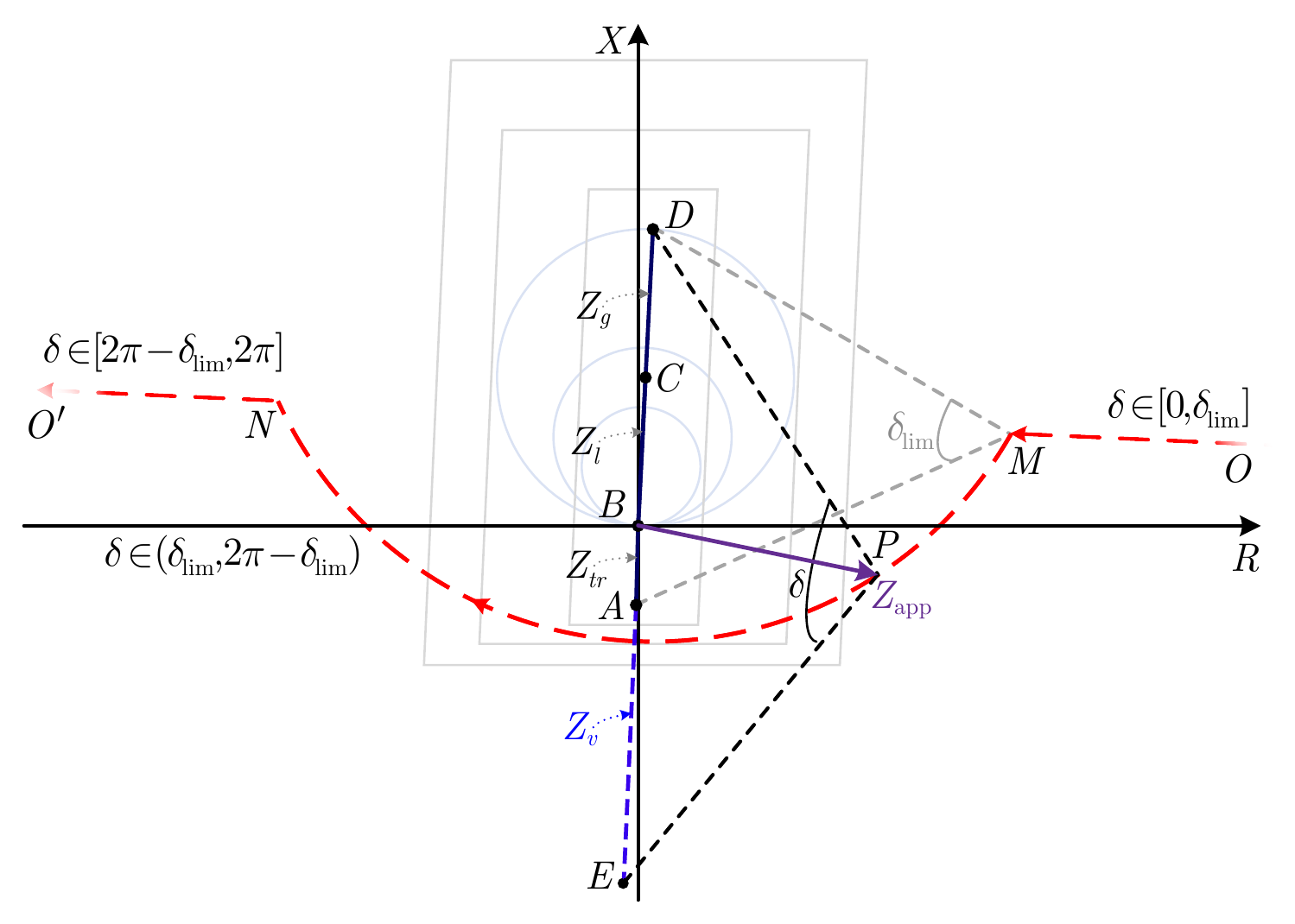}
\caption{Power swing trajectory with the adaptive VI current-limiting strategy}
\label{fig:Saturated Trajectory Adaptive}
\end{figure}

With the adaptive VI current-limiting strategy, the overcurrent is limited to \(I_{\text{max}}\) under all conditions. The VI activates when \(I\) exceeds \(I_{\text{th}}\); however, due to the integrator and PI controller saturation, as shown in Fig.~\ref{fig:Delta V}, the VI remains non-zero and effective only if
\begin{equation}
\label{eq:Enter Saturation Condition}
(i_{sd})^2+(i_{sq})^2 >I_{\text{max}}^2 .
\end{equation}
Therefore, the critical power angle at which the current reaches its maximum allowable value is defined as 
\begin{equation}
\label{eq:Enter Saturation Angle}
\delta_{\text{lim}} := \arccos \left[ \frac{ | \dot{{E}_{\text{ref}}} |^{2}+ | \dot{V_{g}} |^{2}-\left( | Z_{\Sigma} | I_{\text{max}} \right)^2}{2 | \dot{{E}_{\text{ref}}} | | \dot{V_{g}} |} \right].
\end{equation}
The power angle sets for unlimited and limited current are determined by the following conditions: 
\begin{align}
\mathcal{S}_{\text{unlim}} &= \{ \delta \mid \delta \in [0, \delta_{\text{lim}}] \cup [2\pi - \delta_{\text{lim}}, 2\pi] \} 
\label{eq:Unlim_Set} \\
\mathcal{S}_{\text{lim}} &= \{ \delta \mid \delta \in (\delta_{\text{lim}}, 2\pi - \delta_{\text{lim}}) \}. \label{eq:Lim_Set}
\end{align}
The apparent impedance during current limitation is expressed as follows: 
\begin{equation}
\label{eq:Apparent Impedance Saturated Adaptive}
Z_{\text{app}} = \left( Z_{g} + Z_{l} \right) + \frac{| \dot{V_{g}} |}{I_{\text{max}}}\angle \left(-\delta-\theta_{i}\right),
\end{equation}
where \(\theta_{i}\), the phase angle of the limited current vector, is calculated as:
\begin{equation}
\label{eq:Isat Angle}
\theta_{i} = \frac{\pi}{2}-\frac{\delta}{2} - \phi.
\end{equation}
The detailed derivation of~(\ref{eq:Isat Angle}) is provided in Appendix B. Furthermore, the apparent impedance during the current limitation is given by
\begin{equation}
\label{eq:Apparent Impedance Saturated Adaptive as a function of delta}
Z_{\text{app}} = \left(Z_{g} + Z_{l} \right) + \frac{ | \dot{V_{g}} |}{I_{\text{max}}}\angle (-\frac{\delta}{2} - \frac{\pi}{2}+ \phi).
\end{equation}
The full-cycle power swing trajectory with the adaptive VI strategy is shown in Fig.~\ref{fig:Saturated Trajectory Adaptive}. The trajectories \(OM\) and \(NO'\), which represent the unlimited current mode, are identical to those described in Section III.A. When the power angle enters the set \(\mathcal{S}_{\text{lim}}\), the VI is applied with a variable value that changes as \(\delta\) varies. Accordingly, the trajectory moves along an arc centred at \(D\) with a radius of \(DP\), satisfying~(\ref{eq:Apparent Impedance Saturated Adaptive as a function of delta}). The trajectory on the arc shifts with an angular displacement of \(\delta/2\) as \(\delta\) varies. 
Comparing the two current-limiting strategies, the power angles that activate the VI are different, which is reflected in the different positions of point \(M\) on the trajectories. Additionally, when the adaptive VI is activated, the trajectory follows a circular path, whereas the trajectory of the variable VI does not. 
\section{Impact of Power Swing Trajectories on Power Swing Detection}
\label{SectionIV_Impact of Power Swing Trajectories on Power Swing Detection}
In this section, power swing detection in the distance protection scheme is reviewed, followed by an introduction to the APCL and its potential impact on the apparent impedance trajectory.
\subsection{Overview of Power Swing Detection}
\begin{figure}[htbp]
\centering
\includegraphics[width=0.9\columnwidth]{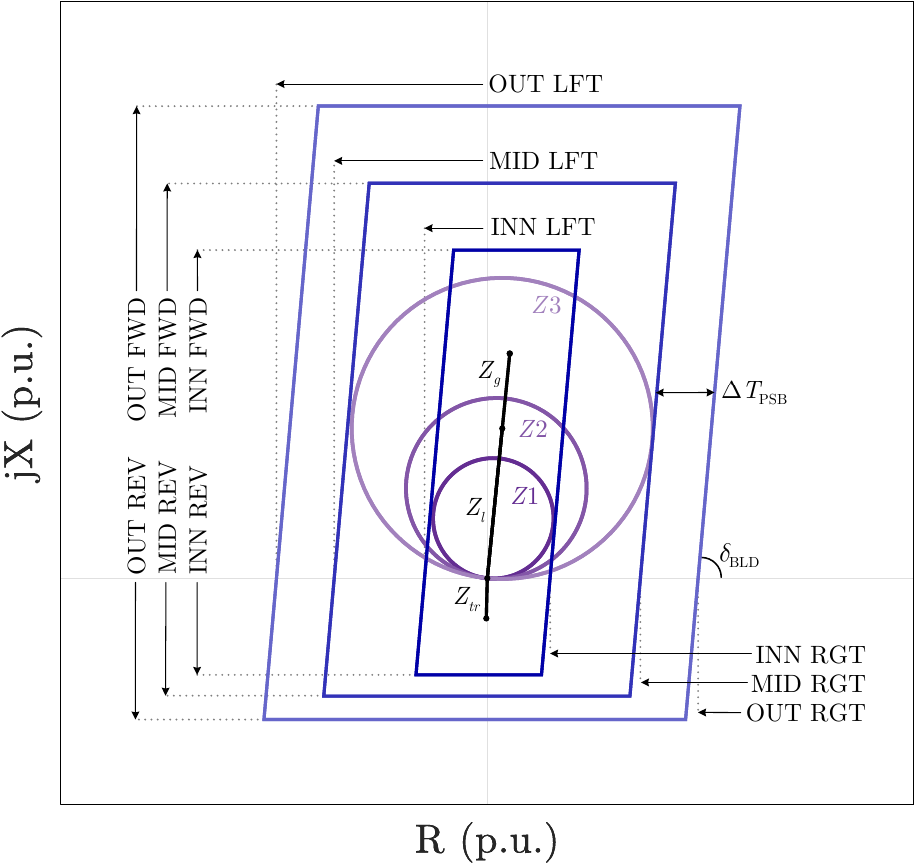}
\caption{Three-zone distance protection characteristics and power swing detection blinder characteristics}
\label{fig:Protection characteristics}
\end{figure}

The conventional power swing detection method responds to the rate of change of apparent impedance to implement two key functions: PSB and OST. This paper applies the three-step resistive blinder scheme shown in Fig.~\ref{fig:Protection characteristics}, which is a more accurate and simpler approach, as outlined in the IEEE Guide~\cite{ieee2016}. The power swing detection settings are configured to block all three distance protection zones during swing events.
\subsubsection{Power Swing Blocking Function}
The primary objective of the PSB function is to differentiate power swings from faults and to block distance relays or other relay elements from operating during power swings. In the three-step resistive blinder scheme, the setting principle relies on the time interval \(\Delta T\), during which the apparent impedance trajectory passes between the outer and middle blinders, to distinguish fault events from power swing events. Faults cause sudden changes in impedance, whereas power swings result in a significantly slower rate of change in the impedance trajectory. The time interval threshold \(\Delta T_{\text{PSB}}\) must be set to guarantee the detection of the fastest power swings~\cite{nerc2013,fischer2012tutorial,psrc2005}. If \(\Delta T > \Delta T_{\text{PSB}}\) is detected, the event is classified as a power swing; therefore, the distance protection zones within the middle blinder are blocked.
\subsubsection{Out-of-Step Tripping Function}
The primary objective of the OST function is to differentiate between stable and unstable power swings and to initiate system separation at specified network locations when the power angle exceeds a predefined threshold \(\delta_{\text{OST}}\), thereby ensuring power system stability and continuity of service. The OST function is implemented through the inner blinder of the three-step resistive blinder scheme. The setting principle of the inner blinder is based on transient stability analysis to ensure that only unstable power swings can travel through it. Once the apparent impedance trajectory travels through the inner blinder, it is identified as an unstable power swing, and a trip command is sent to the circuit breaker to separate the system.
\subsection{Rate of Trajectory Change}
The conventional PSB function relies on the fact that the rate of change of the power swing trajectory in SG-based systems is significantly slower than that of a fault trajectory due to the inherent large inertia of SG rotors. GFM IBRs emulate SGs and synchronise with the power system through APCL, as shown in Fig.~\ref{fig:System Model}(a). Their dynamic behaviour follows the second-order swing equations~\cite{wang2018adaptive}:
\begin{align}
    \label{eq:Swing_Equation1}
    2H \frac{d\omega}{dt} &= P_{0} - P - \frac{1}{D_{p}} \left( \omega - \omega_{0} \right), \\
    \label{eq:Swing_Equation2}
    \frac{d\theta}{dt} &= \omega_{n} \left( \omega - \omega_{0} \right),
\end{align}
where \(P_0\) and \(\omega_0\) represent the setpoints for active power and frequency, respectively; \(H\) and \(D_p\) denote the inertia constant and damping coefficient, respectively; and \(\omega\) and \(\omega_n\) represent the operating frequency and rated frequency, respectively.

\subsection{Factors Affecting the Power Swing Trajectory}
Based on the analysis in Section~\ref{SectionIII_Analysis of Power Swing Trajectories}, the trajectories under the variable and adaptive VI strategies are not straight lines. Instead, the trajectories form curves and may bypass the power swing detection area under specific system configurations. After VI activation, the shape of the trajectory depends on the parameters \(Z_{l}\), \(Z_{g}\), and \(I_{\text{max}}\).

Additionally, since there is no rigid-body inertia in GFM IBRs, their external characteristics are governed by the parameters of the control system. The control parameters \(H\) and \(D_P\) influence the dynamics of power swing, resulting in variations in the impedance trajectory.

\section{Simulation and Case Studies}
\label{SectionV_Simulation and Case Studies}
In this section, after introducing the system configuration and protection settings, simulations are conducted to validate the theoretical trajectories, to investigate the impact of control parameters, and to illustrate the scenarios that cause the power swing detection malfunctions.
\subsection{System Configuration and Protection Setting}
To verify the power swing trajectories under the VI current-limiting strategies, the single-machine GFM IBR grid-connected system shown in Fig.~\ref{fig:System Model}(a) was implemented on the MATLAB/Simulink platform. The red block \(R\) in the figure represents the installed distance relay, which employs the directional three-zone mho characteristics shown in Fig.~\ref{fig:Protection characteristics}. The three-step resistive blinder scheme is employed for power swing detection. The test system parameters, distance protection settings, and power swing detection settings are listed in TABLE~\ref{tab:Parameters and setting}.
\begin{table}[htbp]
\caption{Parameters of the test system, distance protection settings, and power swing detection settings}
\centering
\begin{tabularx}{\columnwidth}{p{1.3cm} X p{1.9cm} } 
\hline
\multicolumn{1}{c}{\textbf{Parameter}} & \multicolumn{1}{c}{\textbf{Description}} & \multicolumn{1}{c}{\textbf{Value}} \\ \hline
\multicolumn{3}{c}{System Configuration}    \\ 
\hline
\(\dot{E_{\text{ref}}}\)   & PCC voltage setpoint  & \(1+j0\) p.u. \\
\(Z_{g}\)          & Impedance of the grid    & 0.6\(\angle84.29 \) p.u.  \\
\(Z_{l}\)          & Impedance of the line         & 0.3\(\angle84.29 \) p.u.  \\
\(Z_{tr}\)         & Impedance of the transformer  & 0.16\(\angle88.57 \) p.u. \\
\(\angle\phi\)     & Impedance angle of \(Z_{\Sigma}\)  & $84.94^{\circ}$           \\
\(I_{\text{max}}\) & Overcurrent limitation        & 1.2 p.u.                  \\
\(I_{\text{th}}\)  & Virtual impedance activation threshold & 1.0 p.u.\\ 
\hline
\multicolumn{3}{c}{Distance Protection Settings}  \\ 
\hline
Z1  & 80\% of \(Z_{l}\)   & 0.48\(\angle84.29 \) p.u. \\
Z2  & 120\% of \(Z_{l}\)  & 0.72\(\angle84.29 \) p.u. \\
Z3  & 200\% of \(Z_{l}\)  & 1.20\(\angle84.29 \) p.u. \\ 
TD1    & Zone1 time delay    & 0 s                    \\
TD2    & Zone2 time delay    & 0.5 s                  \\
TD3    & Zone3 time delay    & 1 s                    \\
\hline
\multicolumn{3}{c}{Power Swing Detection Settings}  \\ 
\hline
OUT RGT    & Outer right blinder    & ~0.84 p.u. \\
OUT LFT    & Outer left blinder     & -0.84 p.u. \\
OUT FWD    & Outer forward reach    & ~1.88 p.u. \\ 
OUT REV    & Outer reverse reach    & -0.56 p.u. \\
MID RGT    & Middle right blinder   & ~0.61 p.u. \\
MID LFT    & Middle left blinder    & -0.61 p.u. \\
MID FWD    & Middle forward reach   & ~1.57 p.u. \\ 
MID REV    & Middle reverse reach   & -0.47 p.u. \\
INN RGT    & Inner right blinder    & ~0.25 p.u. \\ 
INN LFT    & Inner left blinder     & -0.25 p.u. \\
INN FWD    & Inner forward reach    & ~1.31 p.u. \\ 
INN REV    & Inner reverse reach    & -0.39 p.u. \\
\(\delta_{\text{BLD}}\)  & The angles of right and left blinders  & $84.94^{\circ}$  \\ 
\(\Delta T_{\text{PSB}}\)    & Power swing detection time threshold   & 2 cycles \\
\hline
\end{tabularx}
\label{tab:Parameters and setting}
\end{table}

\subsection{Stable Power Swing}
To validate the apparent impedance trajectory derived in Section~\ref{SectionIII_Analysis of Power Swing Trajectories}, Case A is set with a phase jump of $-1.59$ rad occurring at $t=8$s. Case A-I, A-II, and A-III represent the scenarios of no current limitation, the variable VI strategy, and the adaptive VI strategy, respectively.
\begin{figure*}[htbp]
    \centering
    \begin{subfigure}{0.32\textwidth}
        \centering
        \includegraphics[width=\textwidth, clip]{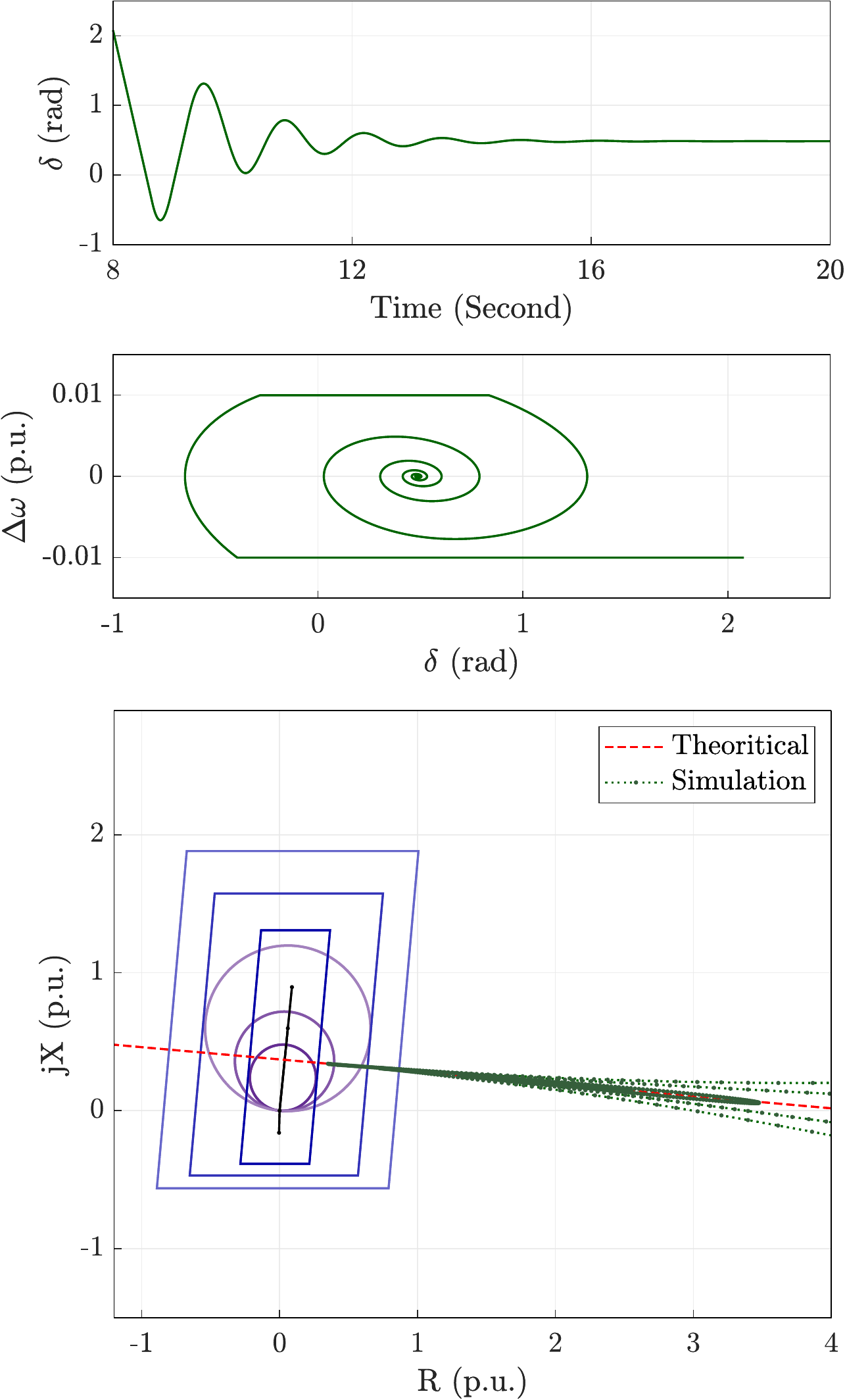} 
        \captionsetup{justification=centering} 
        \caption{}
    \end{subfigure}
    \begin{subfigure}{0.32\textwidth}
        \centering
        \includegraphics[width=\textwidth, clip]{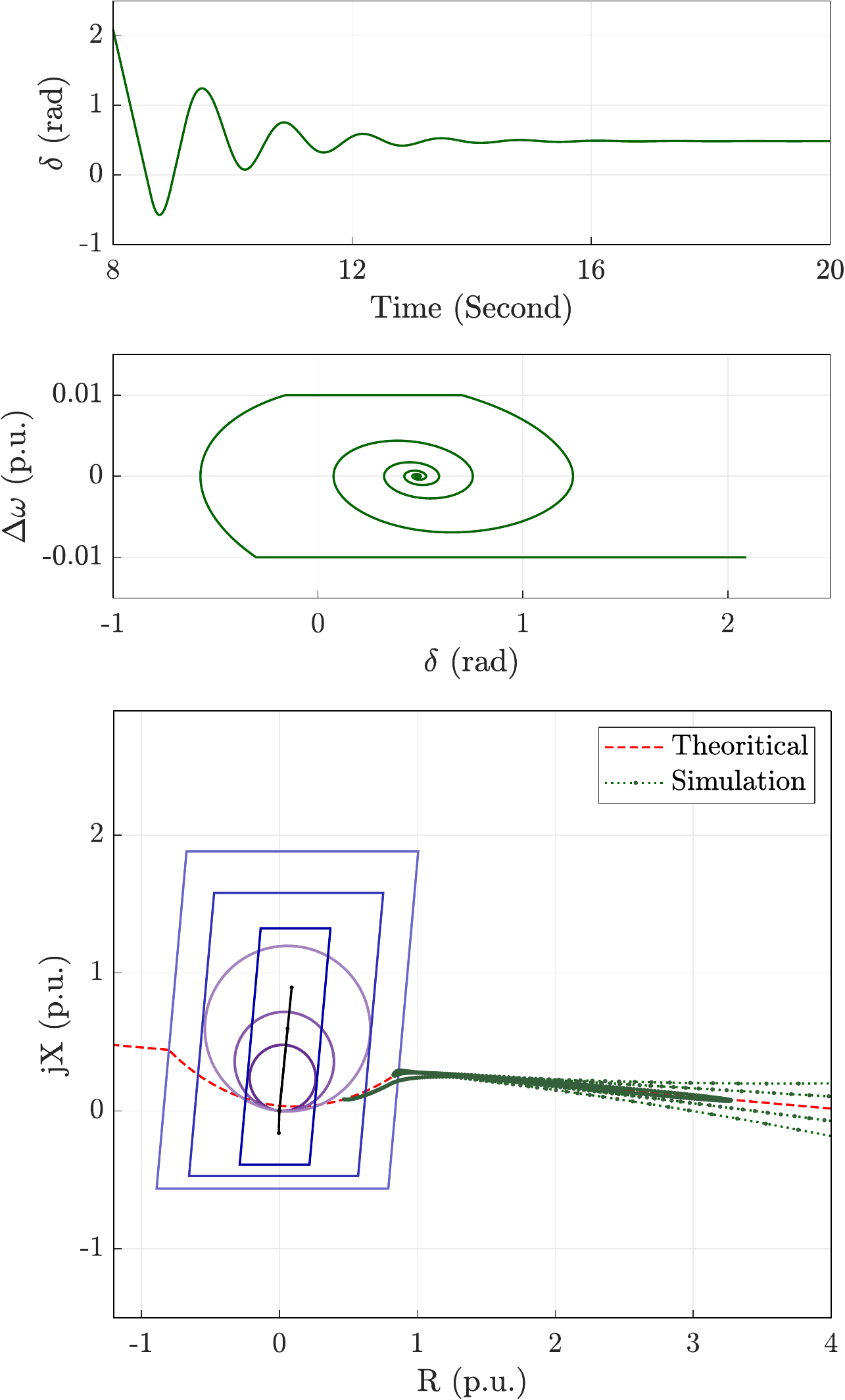} 
        \captionsetup{justification=centering} 
        \caption{}
    \end{subfigure}
    \begin{subfigure}{0.32\textwidth}
        \centering
        \includegraphics[width=\textwidth, clip]{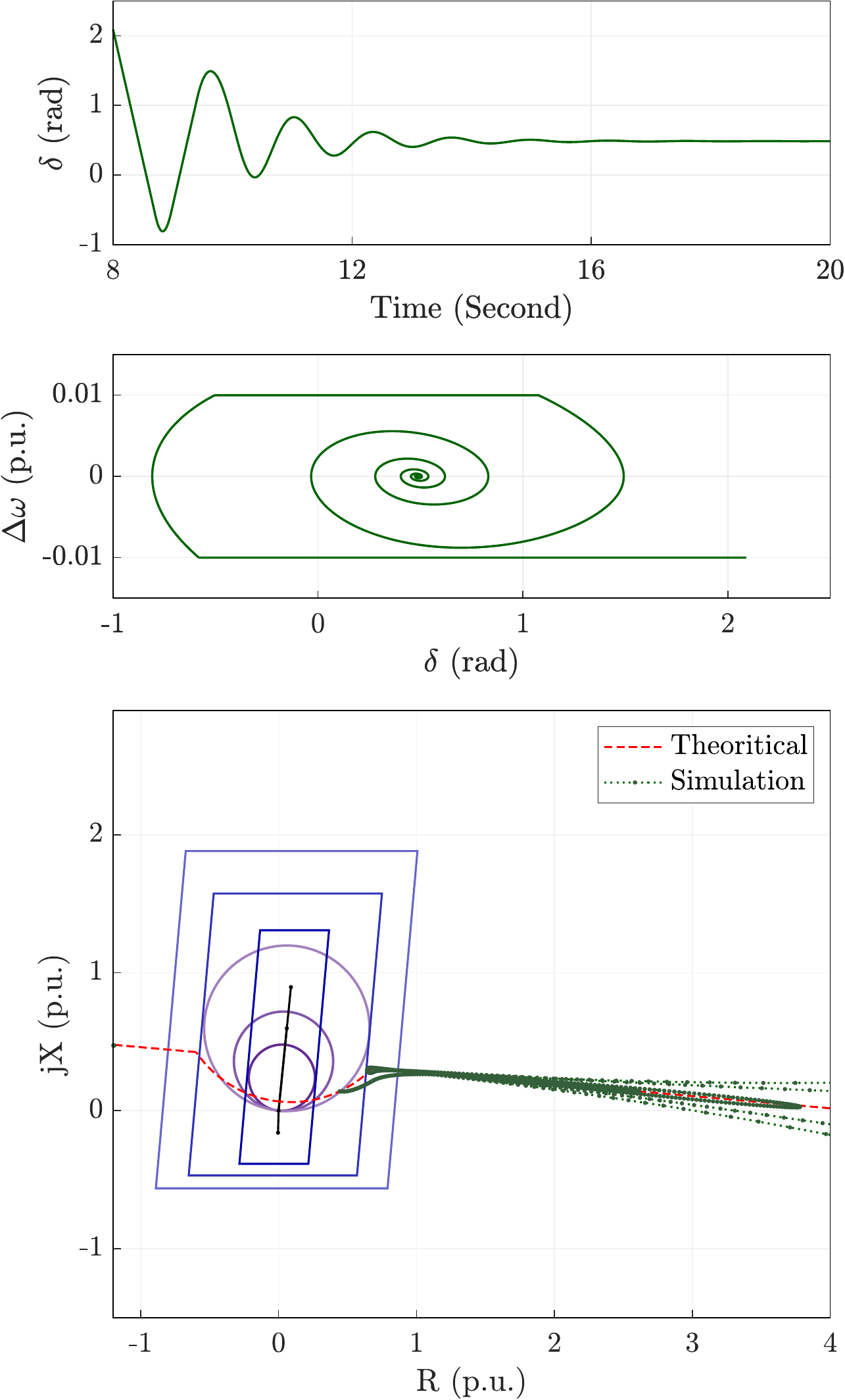} 
        \captionsetup{justification=centering} 
        \caption{}
    \end{subfigure}
    \caption{Results of power angle, phase portrait, and apparent impedance trajectory for Case A. (a) Case A-I: The current is not limited. (b) Case A-II: Variable VI strategy. (c) Case A-III: Adaptive VI strategy.}
    \label{fig:Results for Case A, Case B, and Case C}
\end{figure*}
In Case A, the APCL parameters of three strategies are $D_{p} = 0.05$ and $H = 7$. The setpoint for the active output power is $P_{0}=0.45$~p.u. The other system parameters and protection settings remain consistent with those in TABLE~\ref{tab:Parameters and setting}. The results of Case A are shown in Fig.~\ref{fig:Results for Case A, Case B, and Case C}. 

In Fig.~\ref{fig:Results for Case A, Case B, and Case C}(a), the simulated apparent impedance trajectory in the absence of current limiting closely matches the theoretical trajectory. In Fig.~\ref{fig:Results for Case A, Case B, and Case C}(b), when the VI is not activated, the trajectories align perfectly. However, once the VI is activated, the simulated trajectory exhibits a deviation compared to the theoretical trajectory. This discrepancy arises due to the rapid variation of the power angle, which induces a swift change in overcurrent. Consequently, the effect of the variable VI experiences a slight delay. As shown in Fig.~\ref{fig:Results for Case A, Case B, and Case C}(c), this deviation becomes more pronounced. This is because the adaptive VI is regulated by a proportional-integral (PI) controller, which inherently introduces a slower dynamic response. Additionally, considering the frequency deviation limit in the power system, the APCL's frequency deviation is constrained within ±1\%. As a result, the vertical axis $\Delta \omega$ of the phase portrait curves in all three cases is saturated within the interval \([-0.01, 0.01]\). Furthermore, compared to the power angle \(\delta\) versus time curves in Fig.~\ref{fig:Results for Case A, Case B, and Case C}(b) and Fig.~\ref{fig:Results for Case A, Case B, and Case C}(a), the amplitude of the power swing under the adaptive VI strategy in Fig.~\ref{fig:Results for Case A, Case B, and Case C}(c) is relatively larger. This is also reflected in the apparent impedance trajectory, which spans a longer path.
\subsection{Unstable Power Swing}
To validate the full cycle power swing trajectories, Case B is set. In Case B, a three-phase to ground fault occurs on Line BC at \( t = 4\)s with different current-limiting strategies. 0.25s later, the fault is cleared. 
In Case B, the APCL parameters of three strategies are $D_{p} = 0.05$ and $H = 7$. The setpoint of the active output power is $P_{0}=0.7$~p.u. The other parameters are identical with those in TABLE~\ref{tab:Parameters and setting}. The simulated and theoretical impedance trajectories during the fault recovery process for the absence of the current-limiting strategy, variable VI strategy, and adaptive VI strategy are shown in Fig.~\ref{fig:Results for Case D, Case E, and Case F}(a), (b), and (c), respectively. The simulation trajectories of Case B-I, B-II, and B-III all match the theoretical trajectories, validating the theoretical derivation.
\begin{figure*}[htbp]
    \centering
    \begin{subfigure}{0.32\textwidth}
        \centering
        \includegraphics[width=\textwidth, clip]{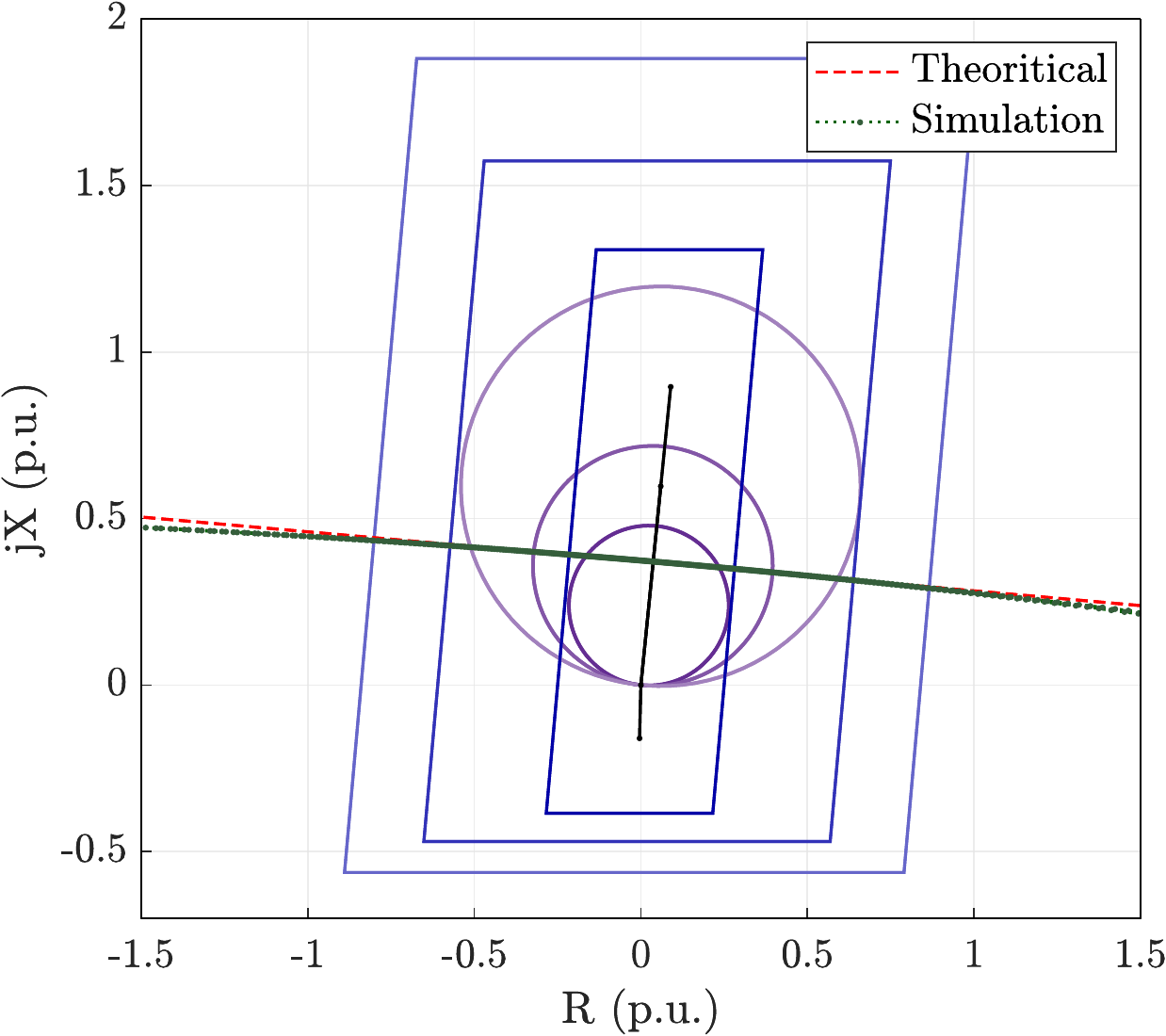} 
        \captionsetup{justification=centering} 
        \caption{}
    \end{subfigure}
    \begin{subfigure}{0.32\textwidth}
        \centering
        \includegraphics[width=\textwidth, clip]{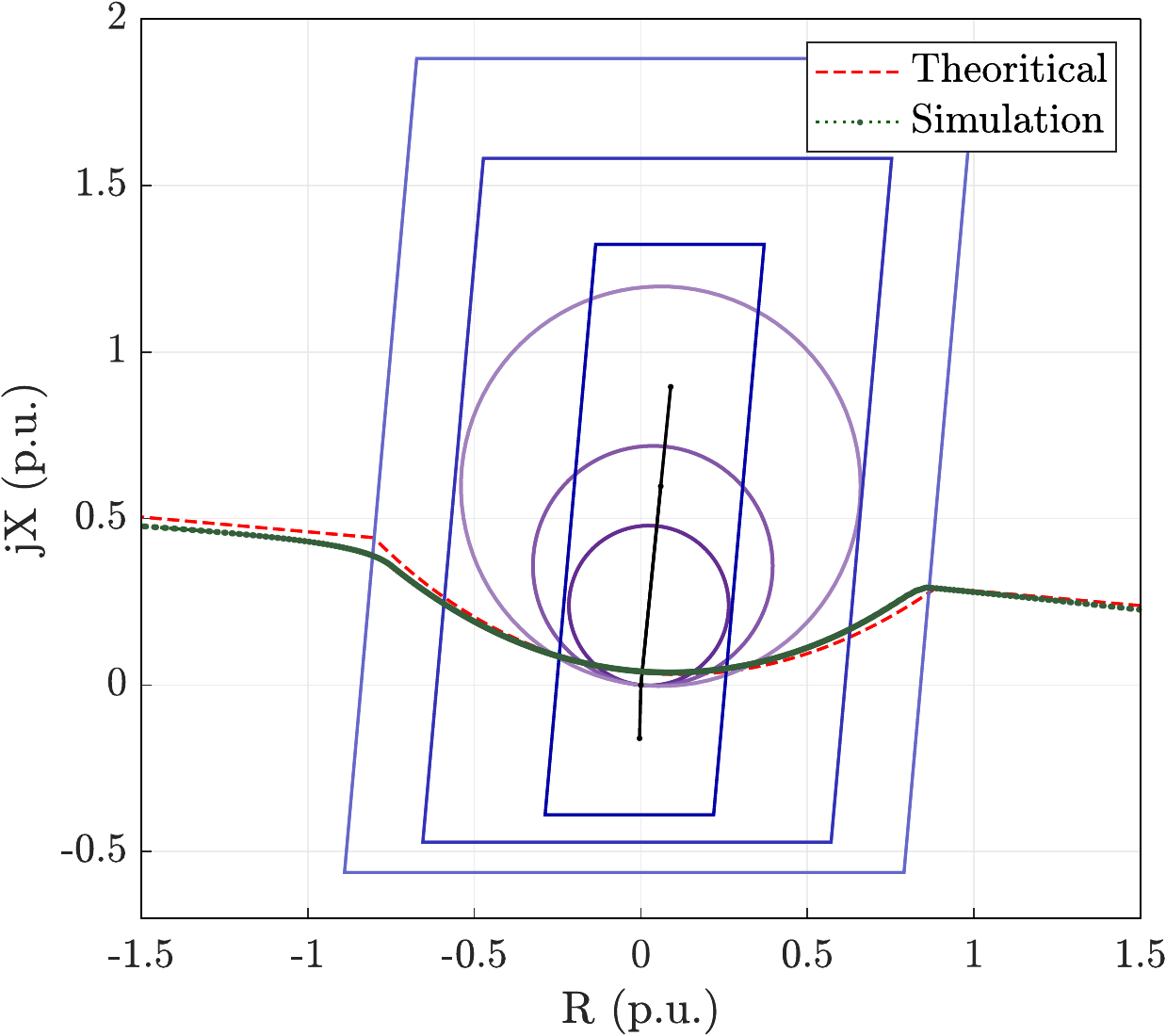} 
        \captionsetup{justification=centering} 
        \caption{}
    \end{subfigure}
    \begin{subfigure}{0.32\textwidth}
        \centering
        \includegraphics[width=\textwidth, clip]{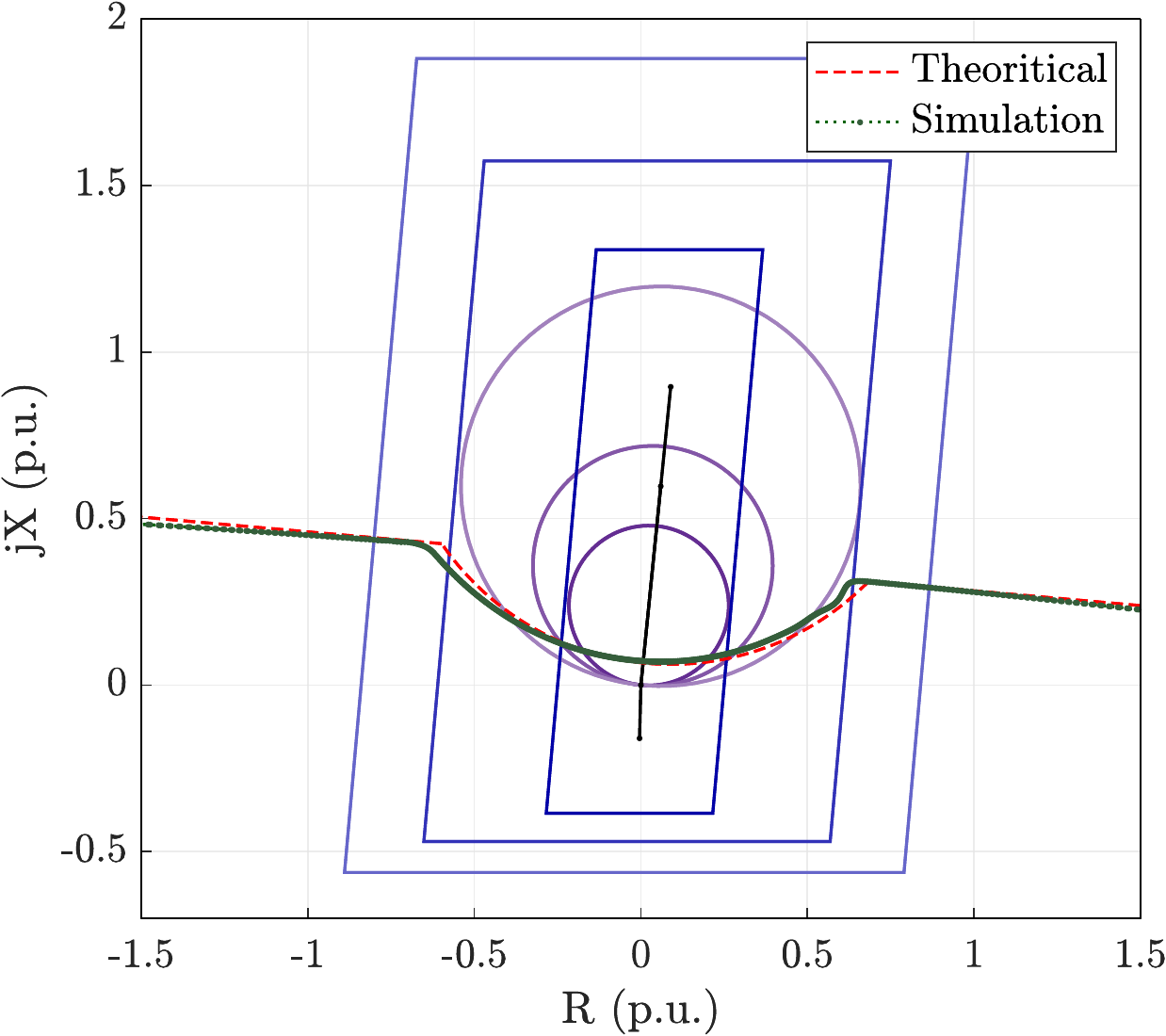} 
        \captionsetup{justification=centering} 
        \caption{}
    \end{subfigure}
    \caption{Results of apparent impedance trajectory for Case B. (a) Case B-I: The current is not limited. (b) Case B-II: Variable VI strategy. (c) Case B-III: Adaptive VI strategy.}
    \label{fig:Results for Case D, Case E, and Case F}
\end{figure*}
\begin{figure*}[h]
    \centering
    \begin{subfigure}{0.32\textwidth}
        \centering
        \includegraphics[width=\textwidth, clip]{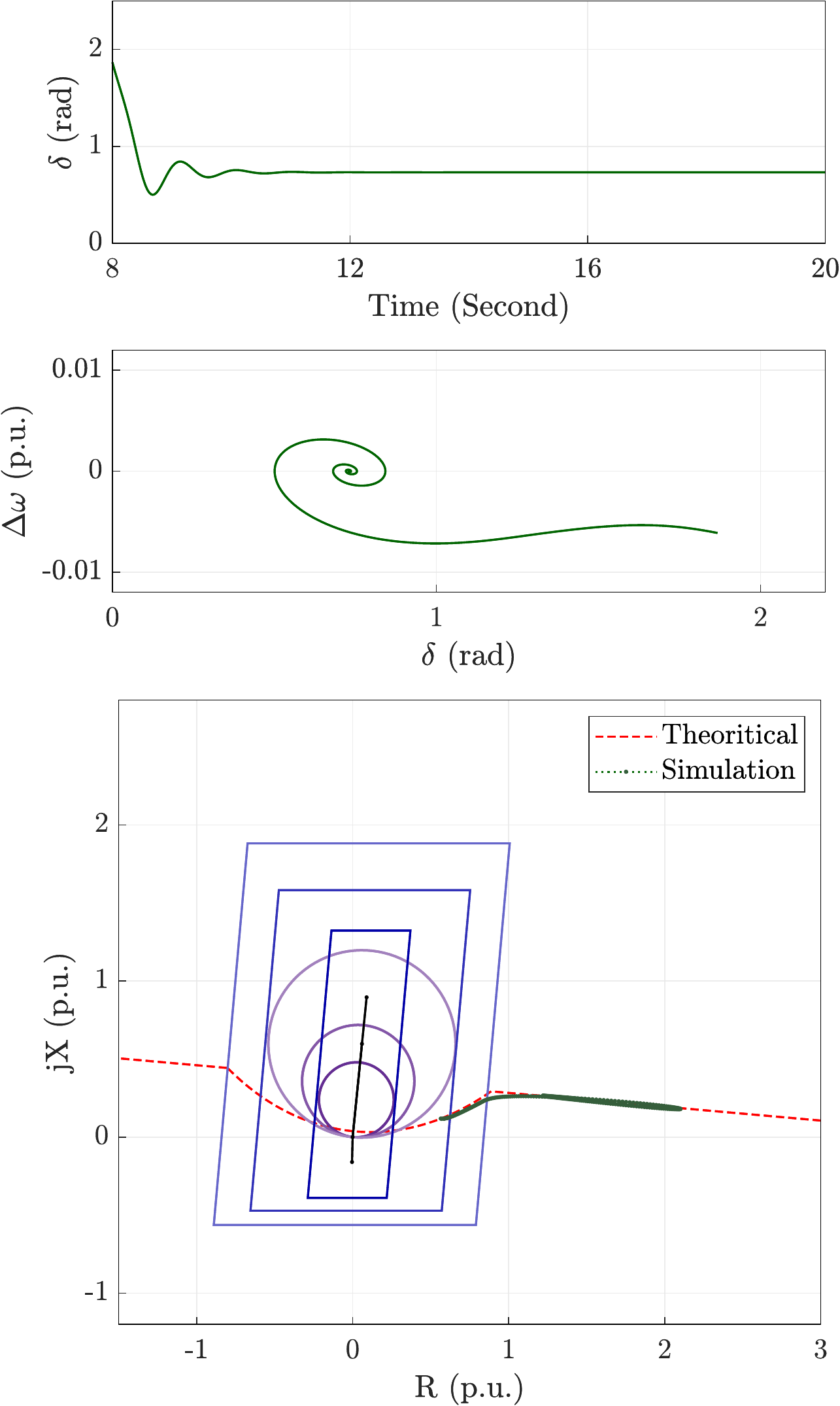} 
        \captionsetup{justification=centering} 
        \caption{}
    \end{subfigure}
    \begin{subfigure}{0.32\textwidth}
        \centering
        \includegraphics[width=\textwidth, clip]{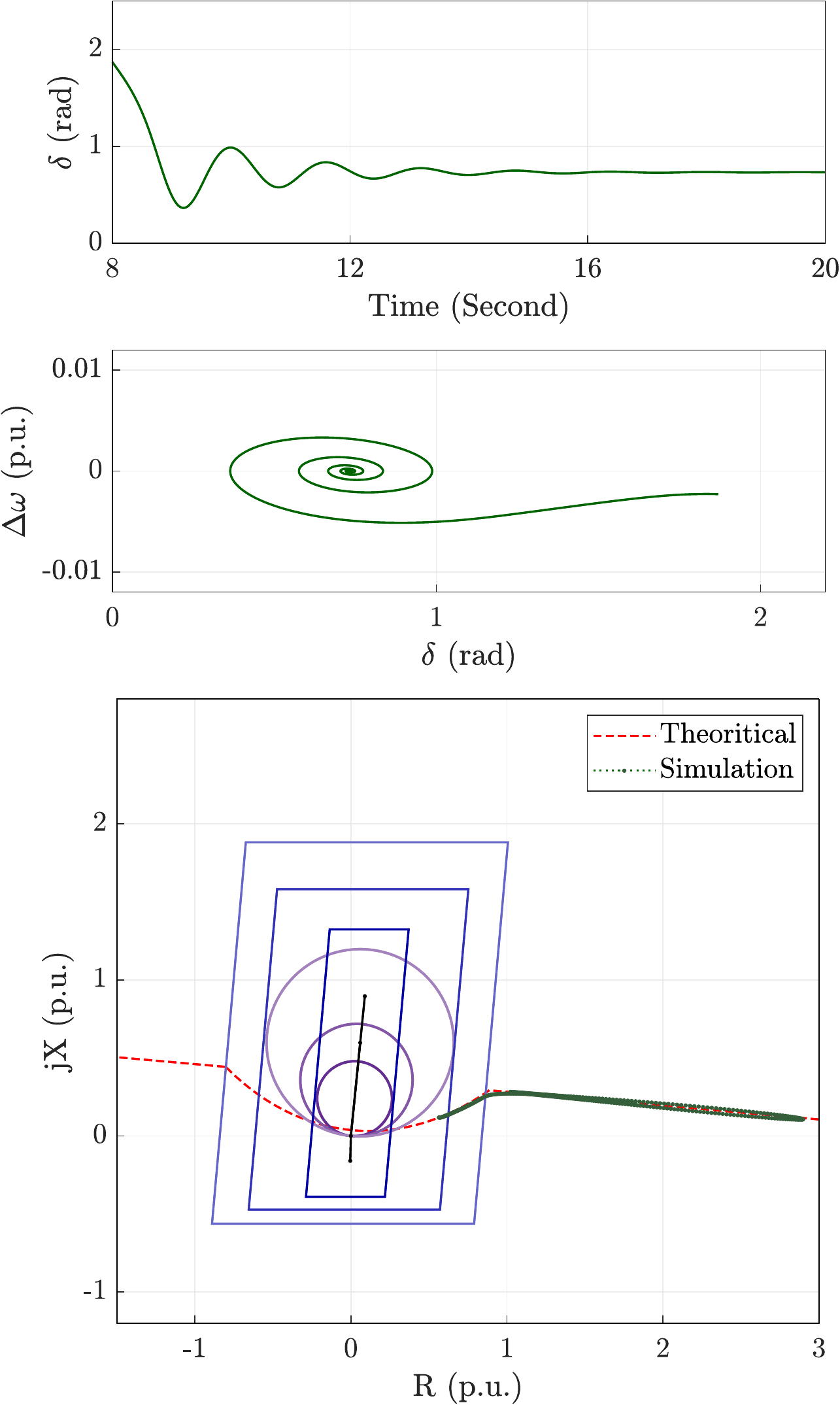} 
        \captionsetup{justification=centering} 
        \caption{}
    \end{subfigure}
    \begin{subfigure}{0.32\textwidth}
        \centering
        \includegraphics[width=\textwidth, clip]{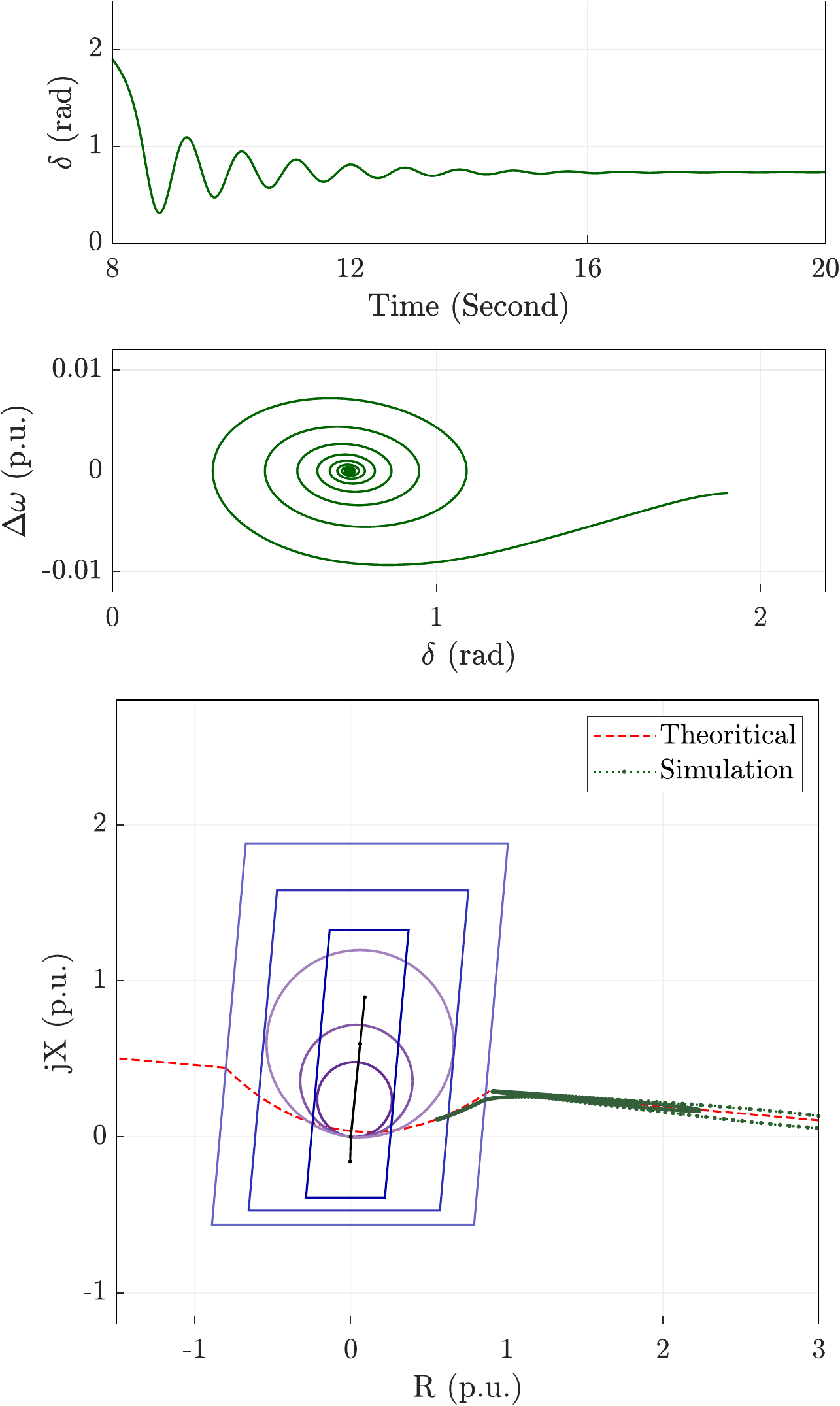} 
        \captionsetup{justification=centering} 
        \caption{}
    \end{subfigure}
    \caption{Results of power angle, phase portrait, and apparent impedance trajectory for Case C. (a) Case C-I: $D_{p} = 0.05$, $H = 3$. (b) Case C-II: $D_{p} = 0.05$, $H = 9$. (c) Case C-III: $D_{p} = 0.15$, $H = 3$.}
    \label{fig:Results for Case G, Case H, and Case I}
\end{figure*}
\subsection{The Impact of Control Parameters}
A GFM IBR does not possess rigid-body inertia, and its external dynamic characteristics are governed by control. Therefore, the control parameters determine power swing performance. Case C is set to evaluate the impact of the APCL parameters \( D_p \) and \( H \). The case settings are as follows.
\begin{itemize}
\item Case C-I: No current limitation; $D_{p} = 0.05$, $H = 3$; $P_{0}=0.65~\text{p.u.}$; at \(t=8\)s, the phase jumps by -1.13 rad.
\item Case C-II: Variable VI strategy; $D_{p} = 0.05$, $H = 9$; $P_{0}=0.65~\text{p.u.}$; at \(t=8\)s, the phase jumps by -1.13 rad.
\item Case C-III: Adaptive VI strategy; $D_{p} = 0.15$, $H = 3$; $P_{0}=0.65~\text{p.u.}$; at \(t=8\)s, the phase jumps by -1.13 rad.
\end{itemize}
The other parameters and settings are same as the data in TABLE~\ref{tab:Parameters and setting}. The results of Case C are shown in Fig.~\ref{fig:Results for Case G, Case H, and Case I}. Comparing Case C-I and Case C-II in Fig.~\ref{fig:Results for Case G, Case H, and Case I}(a) and Fig.~\ref{fig:Results for Case G, Case H, and Case I}(b) respectively, a larger inertia constant \(H\) results in a slower system response to disturbances, reducing the frequency of power swings. 
Therefore, a relatively larger \(H\) is beneficial for the accurate detection of power swings, whereas a relatively smaller \(H\) may compromise the effectiveness of power swing detection. Comparing Case C-I and Case C-III in Fig.~\ref{fig:Results for Case G, Case H, and Case I}(a) and Fig.~\ref{fig:Results for Case G, Case H, and Case I}(c), a larger damping coefficient $D_{p}$ leads to a slower attenuation of oscillations, resulting in a slower dynamic response of the system.
\subsection{Power Swing Detection Malfunction}
\begin{figure}[htbp]
\centering
\includegraphics[width=0.8\columnwidth]{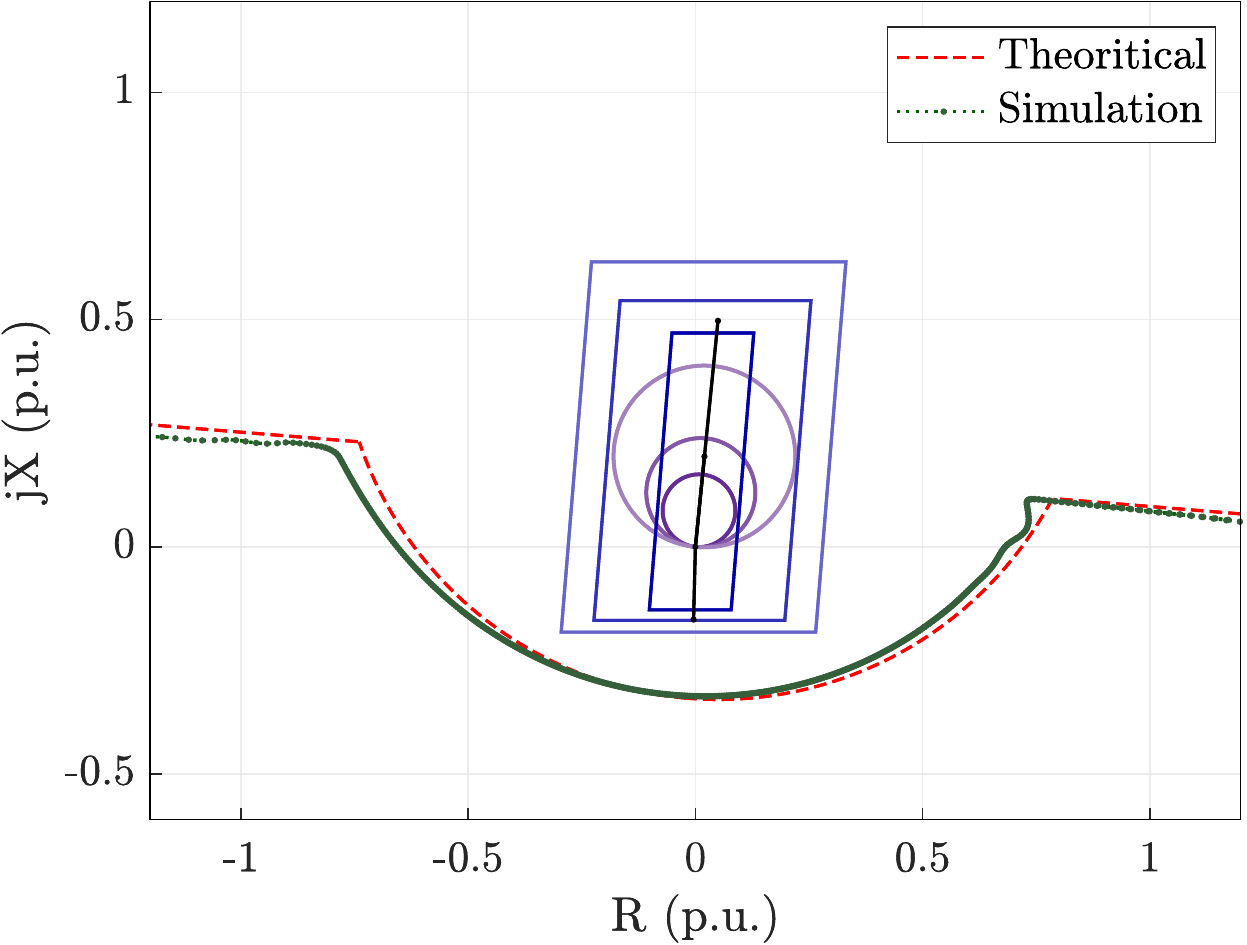}
\caption{Result of apparent impedance trajectory for Case D.}
\label{fig:Result for Case J}
\end{figure}

The power swing trajectory of the IBR with the VI current-limiting strategy is not a straight line. For the variable VI, the trajectory forms a curve, while for the adaptive VI, it follows an arc. Consequently, under certain grid conditions, the power swing trajectories may not pass through the power swing detection area. To illustrate this scenario, Case D is set. Under the adaptive VI strategy, a three-phase to ground fault occurs on Line BC at \( t = 8 \) s; 0.25 s later, the fault is cleared.
The APCL parameters are $D_{p} = 0.05$ and $H = 7$. The setpoint for active output power is $P_{0}=0.7$~p.u. The line impedance is $Z_{l}=0.2$~p.u., and the grid impedance is $Z_{g}=0.3$~p.u. The distance protection settings Z1, Z2, and Z3, as well as the power swing detection blinder settings, are proportionally adjusted according to the change in \(Z_l\), based on the values in TABLE~\ref{tab:Parameters and setting}. After the fault is cleared, the power swing impedance trajectory is shown in Fig.~\ref{fig:Result for Case J}.
In Fig.~\ref{fig:Result for Case J}, the full-cycle power swing trajectory remains outside the power swing detection area. Therefore, the power swing detection functions fail to correctly identify the stable and unstable power swing, thereby compromising their performance.
\subsection{Deterioration of Stability Due to Virtual Impedance}
\begin{figure}[htbp]
    \centering
    \begin{subfigure}{0.88\columnwidth}
        \centering
        \includegraphics[width=\textwidth]{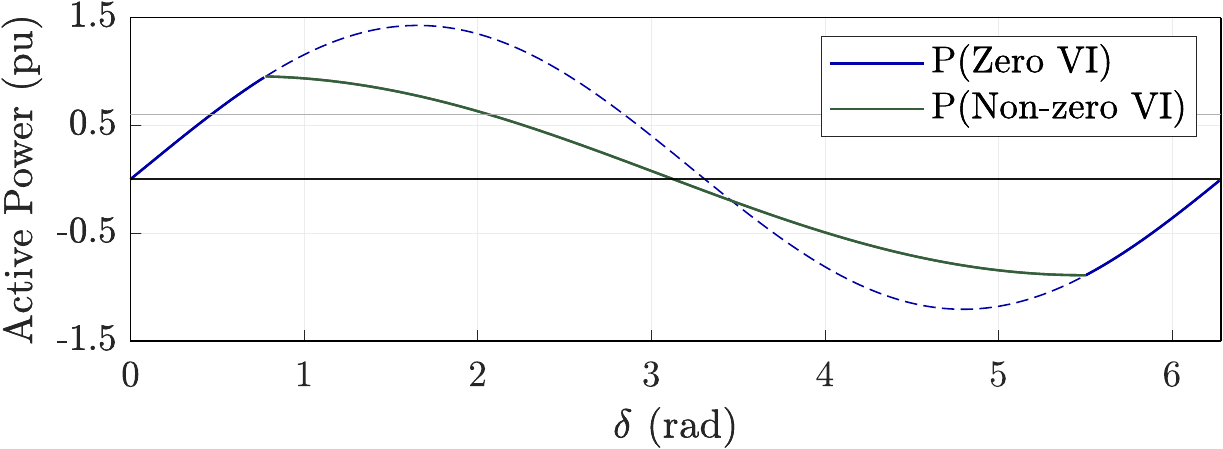}
        \captionsetup{justification=centering} 
        \caption{}
        \label{fig:P-delta_Circular}
    \end{subfigure}
    \begin{subfigure}{0.88\columnwidth}
        \centering
        \includegraphics[width=\textwidth]{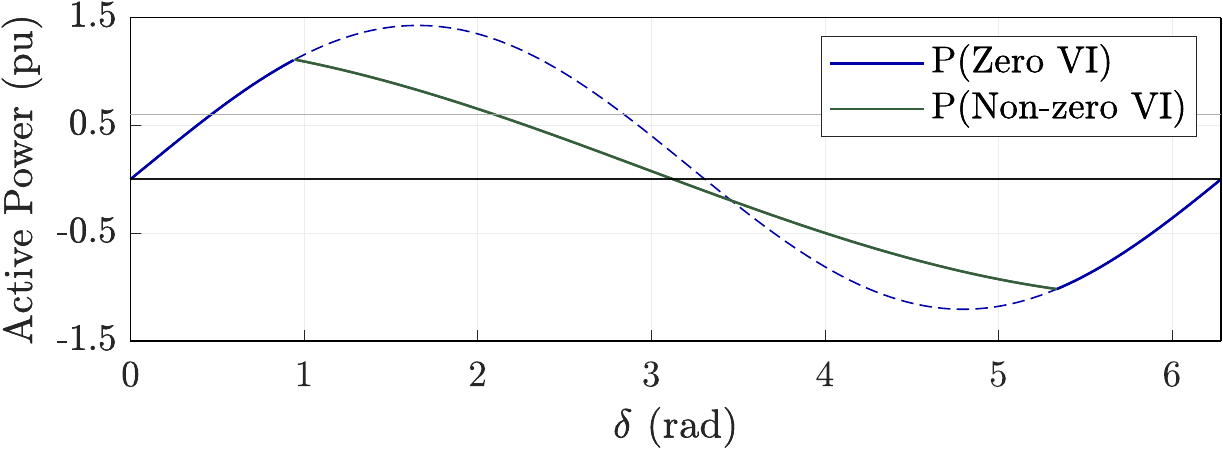}
        \captionsetup{justification=centering}
        \caption{}
        \label{fig:P-delta_D-Priority}
    \end{subfigure}
    \captionsetup{justification=raggedright,singlelinecheck=false} 
    \caption{$P-\delta$ curves for the VI current-limiting strategies. (a) Variable VI. (b) Adaptive VI.}
    \label{fig:P-delta_characteristics}
\end{figure}

When the VI strategy is applied to limit the current, its presence modifies the power angle curve. The output active power when the VI is applied is
\begin{equation}
P_{\text{VI}} = \Re ( \dot{V_{\text{PCC}}} \cdot \dot{I_{\text{VI}}}^* ),
\end{equation}
where
\begin{equation}
\dot{I_{\text{VI}}} = \frac{E_{\text{ref}}\angle0 - V_{g}\angle-\delta}
{(|Z_{\Sigma}|+|Z_{\text{VI}}|)\angle\phi} = I_{\text{VI}} \angle\theta_{I_{\text{VI}}},
\end{equation}
and
\begin{equation}
\dot{V_{\text{PCC}}} = V_g \angle -\delta + |Z_{\Sigma}|\angle \phi \times I_{\text{VI}} \angle\theta_{I_{\text{VI}}}.
\end{equation}
Therefore, the $P-\delta$ curves with the VI strategies are shown in Fig.~\ref{fig:P-delta_characteristics}.
Fig.~\ref{fig:P-delta_characteristics}(a) is the \( P\text{-}\delta \) curve for the variable VI, where the VI is applied when the current exceeds \( I_{\text{th}} \); while Fig.~\ref{fig:P-delta_characteristics}(b) presents the curve for the adaptive VI, where the VI is applied when the current exceeds \( I_{\text{max}} \). Therefore, in Fig.~\ref{fig:P-delta_characteristics}(b), the VI is applied at a relatively larger \( \delta \) compared to Fig.~\ref{fig:P-delta_characteristics}(a). The green curves in the figure reduce the stability margin, which deteriorates the stability of the system. To illustrate this, Case E is set.
\begin{itemize}
\item Case E-I: $D_{p} = 0.05$, $H = 5$; $P_{0}=0.6~\text{p.u.}$; at \(t=8\)s, $\Delta P_{0}=+0.4~\text{p.u.}$.
\item Case E-II: $D_{p} = 0.05$, $H = 5$;$P_{0}=0.6~\text{p.u.}$; at \(t=8\)s, $\Delta P_{0}=+0.5~\text{p.u.}$.
\end{itemize}
Under the same system conditions and the same event, the responses of the three different current-limiting strategies are shown in Fig.~\ref{fig:CaseL,M}. 
\begin{figure}[htbp]
    \centering
    \begin{subfigure}{0.47\columnwidth}
        \centering
        \includegraphics[width=\textwidth]{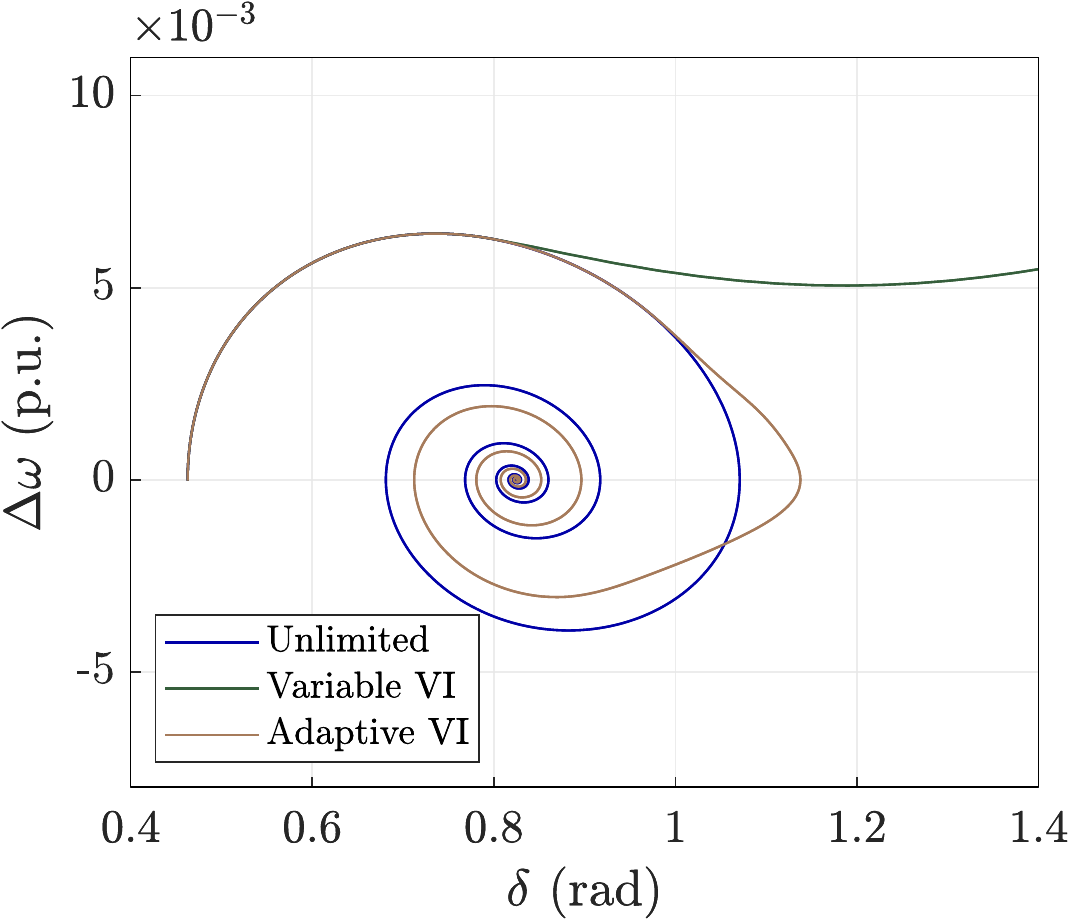}
        \captionsetup{justification=centering} 
        \caption{} 
        \label{fig:Circular_Priority}
    \end{subfigure}
    \begin{subfigure}{0.47\columnwidth}
        \centering
        \includegraphics[width=\textwidth]{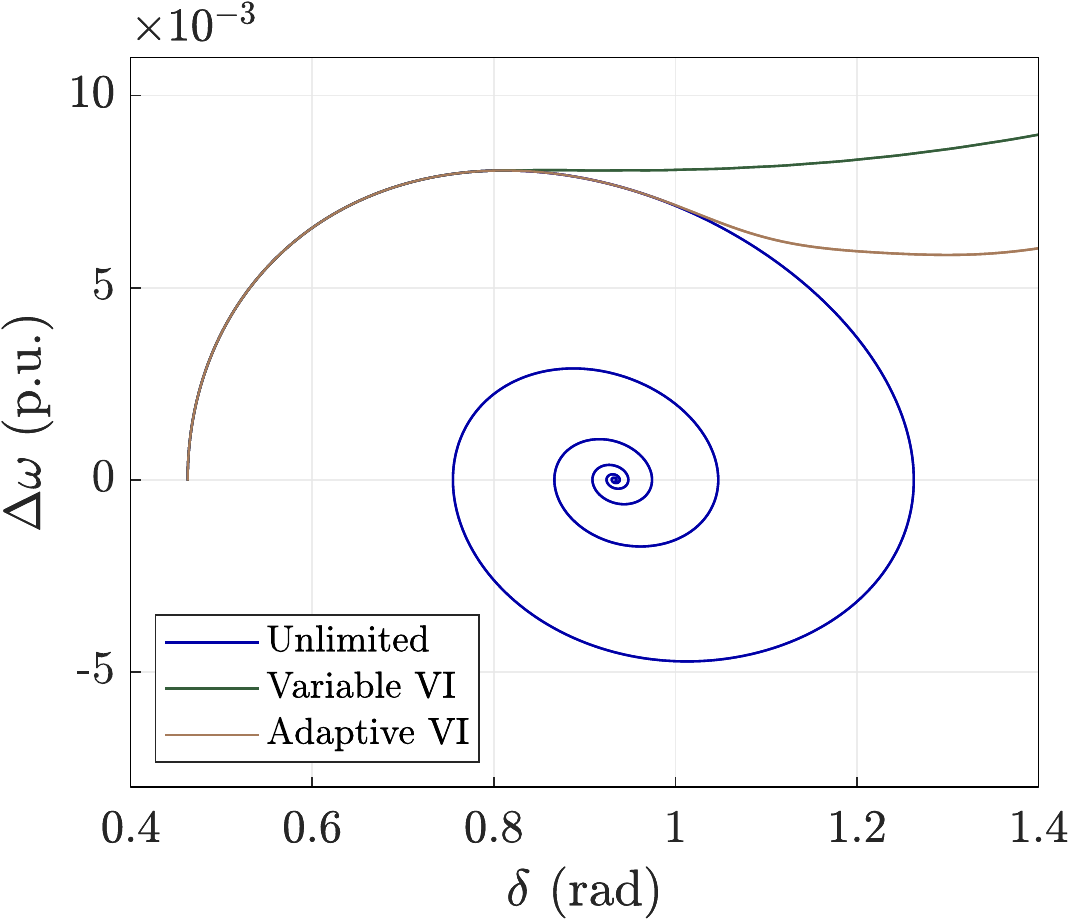}
        \captionsetup{justification=centering} 
        \caption{} 
        \label{fig:D_Axis_Priority}
    \end{subfigure}
    \captionsetup{justification=raggedright,singlelinecheck=false} 
    \caption{The phase portraits with different current-limiting strategies of Case E (a) Case E-I. (b) Case E-II.}
    \label{fig:CaseL,M}
\end{figure}
In Case E-I, both the unlimited current scenario and the adaptive VI strategy lead the system to a stable equilibrium point, whereas the variable VI strategy results in instability. In Case E-II, two of the VI strategies become unstable, and only the unlimited current scenario returns to a stable state. Case E demonstrates that both the variable and adaptive VI strategies may deteriorate the stability of the system; however, the variable VI strategy has a more severe impact.
\section{Conclusion}
\label{SectionVI_Conclusion}
This paper theoretically analyses the apparent impedance trajectories of GFM IBR under the variable VI and adaptive VI current-limiting strategies. The trajectories under these two strategies exhibit fundamental differences compared with those of an SG and a GFM IBR without current limiting. The trajectories are validated through simulations built on the MATLAB/Simulink platform under both stable and unstable power swings. Since the dynamic characteristics of GFM IBR are control-based, the impact of APCL control parameters on the trajectories is analysed. Furthermore, simulation results indicate that the new trajectory characteristics under the VI strategies may cause conventional power swing detection functions to fail. Additionally, the results show that the introduction of VI deteriorats the transient stability of the system.

{\appendix[A. The Voltage Drop Across the Virtual Impedance]
The voltage drop vector across the VI under the variable VI strategy is 
\begin{equation}
\label{}
Z_{\text{VI}}\dot{I_{\text{max}}}=V_{d\text{VI}}+jV_{q\text{VI}}.
\end{equation} 
By substituting \(Z_{\text{VI}} = R_{\text{VI}}+jX_{\text{VI}}\) with~(\ref{eq:VI Rv}) and~(\ref{eq:VI Xv}) respectively, one will reach
\begin{align}
\label{}
& [  k_{\text{VI}} ( I_{\text{max}} - I_{\text{th}} ) 
+ j \alpha_{\text{VI}} k_{\text{VI}} ( I_{\text{max}} - I_{\text{th}} )  ] 
\notag \\
& \quad \times ( i_{sd} + j i_{sq} ) 
= V_{d\text{VI}} + j V_{q\text{VI}}.
\end{align}
Thus, the magnitude of the voltage drop is
\begin{align}
\label{}
\Delta V &= \sqrt{V_{d\text{VI}}^2 + V_{q\text{VI}}^2} \notag \\
&= \sqrt{
    \left\{ 
        \left[ 
            k_{\text{VI}} ( I_{\text{max}} - I_{\text{th}} ) i_{sd} 
            - \alpha_{\text{VI}} k_{\text{VI}} ( I_{\text{max}} - I_{\text{th}} ) i_{sq} 
        \right]^2 
    \right. 
} \notag \\
&\quad + 
\left. 
    \left[ 
        k_{\text{VI}} ( I_{\text{max}} - I_{\text{th}} ) i_{sq} 
        + \alpha_{\text{VI}} k_{\text{VI}} ( I_{\text{max}} - I_{\text{th}} ) i_{sd} 
    \right]^2 
\right\}, 
\end{align}
and is equal to
\begin{equation}
\label{}
\Delta V = k_{\text{VI}}( I_{\text{max}} - I_{\text{th}})I_{\text{max}}\sqrt{1+\alpha_{\text{VI}}^2}.
\end{equation} 
}
 
{\appendices
\section*{B. Current Argument Under Adaptive Virtual Impedance Strategy}
According to the Kirkhoff's Voltage Law, the voltage-current relationship in the equivalent circuit shown in Fig.~\ref{fig:Single Machine Grid Connected Model with VI} is calculated as
\begin{equation}
\label{}
E_{\text{ref}}\angle0-V_{g}\angle-\delta = I_{\text{max}}\angle \theta_{i}( |Z_{\text{VI}}|+|Z_{\Sigma}| )\angle\phi,
\end{equation} 
which is formally equal to
\begin{equation}
\label{}
(e_{d}^{\text{ref}}-V_{g}\cos\delta ) +j V_{g}\sin\delta = I_{\text{max}}\angle \theta_{i}( |Z_{\text{VI}}|+|Z_{\Sigma}| )\angle\phi.
\end{equation} 
Considering that the phase angles on both sides of the equation are equal, therefore,
\begin{equation}
\label{}
\arctan\left(\frac{V_{g}\sin\delta}{e_{d}^{\text{ref}}-V_{g}\cos\delta}\right) = \theta_{i}+\phi.
\end{equation}
Assuming \(|\dot{V}_{g} | = |\dot{E_{\text{ref}}} |\), the following trigonometric relationship holds
\begin{equation}
\label{}
\arctan\left(\frac{\sin\delta}{1-\cos\delta}\right) = \frac{\pi}{2}-\frac{\delta}{2}.
\end{equation}
Therefore,
\begin{equation}
\label{}
\theta_{i} = \frac{\pi}{2}-\frac{\delta}{2}-\phi.
\end{equation}
\section*{}
}



\footnotesize
\bibliographystyle{IEEEtran}
\bibliography{IEEEabrv,main}

\clearpage

\ifCLASSOPTIONcaptionsoff
  \newpage
\fi
\end{document}


\captionsetup{font={small}}

\title{Supplementary Information
}

\markboth{Supplementary Information}%
{Shell \MakeLowercase{\textit{et al.}}: Bare Demo of IEEEtran.cls for IEEE Journals}

\maketitle

\IEEEpeerreviewmaketitle

\appendices
\section{Separating }
\label{Appendix_deduction}
The 
\section{Training }
The 

\footnotesize

\ifCLASSOPTIONcaptionsoff
  \newpage
\fi